\documentclass{aa}

\usepackage{txfonts}
\usepackage{amsmath,amssymb,amsfonts}
\usepackage{mathrsfs}

\usepackage{graphicx}
\usepackage{subcaption}
\usepackage{float}
\usepackage{rotating}

\usepackage{longtable}
\usepackage{threeparttable}
\usepackage{booktabs}
\usepackage{multirow}

\usepackage[version=4]{mhchem}

\usepackage{algorithm}
\usepackage{algorithmicx}
\usepackage{algpseudocode}
\usepackage{listings}

\usepackage[title]{appendix}

\usepackage{xcolor}
\usepackage{textcomp}
\usepackage{scalerel}
\usepackage{placeins}
\usepackage{changepage}
\usepackage{soul}
\usepackage{etoolbox}
\usepackage{chngcntr}

\usepackage[
    colorlinks=true,
    linkcolor=blue,
    citecolor=blue,
    urlcolor=blue
]{hyperref}

\newcommand{\kmps}{km\,s$^{-1}$}
\newcommand{\msun}{M$_\odot$}
\newcommand{\mpy}{M$_\odot$\,yr$^{-1}$}
\newcommand{\mpcinv}{Mpc$^{-1}$}
\newcommand{\pow}[1]{10^{#1}}
\newcommand{\pcc}{cm$^{-3}$}

\newcommand{\uncovgal}{{\tt UNCOVER 3686\,}}

\newcommand{\oii}{\textsc{O ii}}
\newcommand{\oiii}{\textsc{O iii}}
\newcommand{\neiii}{Ne\,\textsc{iii}}

\setstcolor{cyan}

\defcitealias{dekel2023efficient}{D23}

\newlength{\imagew}
\newlength{\imageh}
\newlength{\legendw}
\newlength{\legendh}
\newlength{\legendx}
\newlength{\legendy}

\newcommand{\graphicswithlegend}[6]{%
  \setlength{\imagew}{#1}
  \settoheight{\imageh}{\includegraphics[width=\imagew]{#2}}

  \setlength{\legendw}{#3\imagew}
  \settoheight{\legendh}{\includegraphics[width=\legendw]{#4}}

  \setlength{\legendx}{\imagew}
  \addtolength{\legendx}{-\legendw}
  \addtolength{\legendx}{-#5\imagew}

  \setlength{\legendy}{\imageh}
  \addtolength{\legendy}{-\legendh}
  \addtolength{\legendy}{-#6\imageh}

  \includegraphics[width=\imagew]{#2}%
  \llap{%
    \hspace{-\the\legendx}
    \raisebox{\legendy}{\includegraphics[width=\legendw]{#4}}
    \hspace{\the\legendx}
  }
}

\counterwithin{figure}{section}
\counterwithin{table}{section}

\begin{document}

\title{A Massive Galaxy at the Edge of Feedback-Free Efficiency} 

\author{
K. Aditya\inst{1,2,3}\corrauth{kaditya.astro@gmail.com}
\and
Kartick C. Sarkar\inst{3}\email{kcsarkar@rri.res.in}
}

\institute{Instituto de Estudios Astrofísicos, Facultad de Ingeniería y Ciencias, Universidad Diego Portales, Av. Ejército 441, Santiago, Chile
\and
Millennium Nucleus for Galaxies, Concepción, Chile
\and
Raman Research Institute, C. V. Raman Avenue, Sadashivanagar, Bengaluru, 560080, India}

\date{Received XX XX 2026 / Accepted XX XX 2026}

\abstract
{The efficiency with which galaxies convert their available baryonic reservoir into stars sets a fundamental ceiling on stellar mass assembly in the early Universe and encodes the cumulative effect of stellar feedback. in this paper, we report the measurement of star-formation efficiency (SFE) of a photometrically and spectroscopically vetted reference sample at $z=9-10$, out of which 31 are spectroscopically confirmed. Among them, we highlight a spectroscopically confirmed galaxy, \texttt{UNCOVER 3686} at $z = 9.31$, which has a stellar mass $M_\star = 10^{9.55}$ M$_\odot$ and physical properties comparable to the predictions of the feedback-free starburst (FFB) scenario. We use this galaxy as an anchor for the first direct observational test of whether galaxy at cosmic dawn with physical properties predicted in the FFB theory reach the maximum baryon conversion efficiencies predicted by the feedback-free starburst scenario. We find the SFE for this galaxy to lie between $\approx 20\%$ and $60\%$, a factor of 2 to 6 above empirical model predictions. The compact morphology ($R_e = 0.45$ kpc), young stellar age ($\sim 160$ Myr), low metallicity $(Z_\star/Z_\odot \approx 0.2)$, and inferred gas density $(n_{\rm gas} \sim 3\times 10^{3}$ cm$^{-3}$) of this galaxy are consistent with feedback-free (FFB) galaxy formation conditions. We conclude that \uncovgal is an excellent candidate for an FFB galaxy in which the global star formation efficiency approaches the theoretical limits due to weak stellar feedback.} 


\maketitle
\nolinenumbers
\section{Introduction}\label{sec1}
The James Webb Space Telescope (JWST) has uncovered a substantial population of massive galaxies with stellar masses $M_{\star} \gtrsim 10^{9}\,\mathrm{M_{\odot}}$ within just
$500$--$600$ million years after the Big Bang \citep{2023NatAs...7..622C, 2023ApJ...951L..22A, 2023MNRAS.523.1009B, 2024ApJ...973....8H, 2023Natur.618..480R, 2023Sci...380..416W, 2023A&A...677A.115C, 2024NatAs...8..657B, 2018Natur.557..392H, 2023MNRAS.526.1657T, 2023ApJ...955..130F, 2023ApJS..269...33N, 2015ApJ...810L..12Z, 2024ApJ...962...24S, naidu2022two}. Some of these galaxies appear to have assembled stellar masses as high as $10^{10}\,\mathrm{M_{\odot}}$ within the first $\sim$500 million years after the Big Bang \citep{labbe2023population}, contrary to our traditional understanding of the efficiency with which stars form in galaxies. These observations have prompted a fresh examination of galaxy formation models within the $\Lambda$CDM framework \citep{boylan2023stress}. However, further NIRSpec follow-up observations have confirmed that several of these photometrically selected systems are in fact at redshifts $z \approx 4$-$6$ \citep{kocevski2023hidden, harikane2023comprehensive, arrabal2023confirmation, 2023ApJ...951L..22A}, rather than $z \gtrsim 9$ as initially inferred from photometric redshift estimates and some identified as AGN-dominated little red dots rather than purely stellar systems \citep{kocevski2023hidden,kokorev2024census}.  This is consistent with spectroscopic studies reporting massive galaxies ($M_{\star} = 10^{10}$-$10^{11}\,\mathrm{M_{\odot}}$) at $z \approx 4$-$5$ \citep{glazebrook2024massive, carnall2023massive, de2025efficient}, underscoring the need for spectroscopically confirmed samples with robust stellar mass estimates.
 
The maximum stellar mass a galaxy can attain is fundamentally limited by how efficiently the available baryonic reservoir within its dark matter halo is converted into stars. Observation-based phenomenological models, including abundance matching and halo occupation distribution frameworks, predict a maximum star formation efficiency of $\varepsilon \approx 0.1$ \citep{2013MNRAS.428.3121M, 2018MNRAS.477.1822M, 2018ApJ...868...92T,2018ARA&A..56..435W}. It has been shown that at least two galaxies from the red-galaxy population reported in \cite{labbe2023population} at $z = 7$-$10$ require extraordinarily high efficiencies, $\varepsilon > 0.5$ \citep{boylan2023stress}. However, subsequent NIRSpec follow-up confirmed that one of these galaxies lies at $z \approx 5.6$ \citep{kocevski2023hidden, 2023ApJ...951L..22A}. At this redshift, the inferred efficiency falls within the range expected from standard phenomenological models. For example, spectroscopic studies of massive galaxies by \cite{xiao2024accelerated} find baryon conversion efficiencies of $\varepsilon \approx 50\%$, for dusty star forming galaxies (DFSGs) $z \sim 5$-$6$ in the \texttt{FRESCO} survey. This high efficiency across $4<z<9$ is also corroborated by stellar mass function measurements using photometrically selected galaxies \citep{weibel2024galaxy, chworowsky2024evidence}.

These observations raise a fundamental question: can a fraction of the high-redshift galaxy population form stars at near-maximal efficiency, approaching the physical ceiling set by the total baryonic budget of their host halos? Such extreme efficiencies are expected to arise naturally in the feedback-free starburst (FFB) scenario \citep{dekel2023efficient,li2024feedback}. In this picture, the gas accreted into massive halos at $z \sim 10$ reaches sufficiently high densities ($n_{\mathrm{gas}} \gtrsim 10^{3}\,\mathrm{cm^{-3}}$) such that free-fall timescales fall below $\sim 1$ Myr, shorter than the timescale on which stellar winds and supernova feedback develop in a dense cloud. Under these conditions, star formation is unimpeded, enabling near-complete conversion of accreted baryons into stars. Massive galaxies in $\sim$$10^{11}\,\mathrm{M_{\odot}}$ halos at $z \sim 10$ are therefore expected to achieve $\varepsilon \sim 1$, driven by high cosmological gas densities and low metallicities characteristic of that epoch. At later times, global efficiencies of $\varepsilon \sim 0.01-0.1$ arise as a direct outcome of feedback from supernovae and Active Galactic Nuclei (AGN), which suppress star formation by heating or ejecting gas from star-forming regions. Several complementary frameworks have been proposed 
to explain efficient star formation at $z>9$,  including modifications to stellar feedback prescriptions, bursty star formation histories, 
and variable dust attenuation; for a review see \cite{adamo2025first, somerville2026galaxy}. Among these, the FFB scenario is particularly 
amenable to observational tests because it makes a specific, quantitative prediction: galaxies satisfying the physical threshold conditions of 
halo mass, gas density, and metallicity should attain near maximal baryon conversion efficiencies $\varepsilon= 0.2$-$1$ 
\citep{li2024feedback}. The \cite{dekel2023efficient} framework is particularly well suited to describe galaxy formation at $z \geq 9$ and above, 
where the halo densities required for feedback-free star formation are most readily achieved, and predicts baryon 
conversion efficiencies as an explicit function of redshift \citep{li2024feedback}.
Here we perform the first direct observational test of this prediction, measuring the global baryon conversion efficiency 
of \uncovgal\ ($z = 9.31$) through rank-ordered abundance matching against a reference sample of 142 galaxies including 31 spectroscopically 
confirmed galaxies between $9 < z < 10$ spanning $0.2~\mathrm{deg}^2$, and comparing the result directly with the quantitative predictions from FFB theory 
\citep{dekel2023efficient, li2024feedback}. We characterize the physical properties of \uncovgal in Section~\ref{sec2} and assess whether 
they satisfy the conditions predicted for feedback-free star formation, then present the first quantitative measurement of its baryon 
conversion efficiency in Section~\ref{sec:results} and compare it directly to the theoretical prediction from FFB. Finally, we will draw our conclusions in Section \ref{sec:conclusions}

\section{Characterizing \uncovgal}\label{sec2}

\subsection{Data} \label{subsec:Data}
We identify the galaxy from the publicly available JWST data from the \textsc{ULTRADEEP}, \textsc{NIRSPEC}, and \textsc{NIRCAM OBSERVATIONS BEFORE THE EPOCH OF REIONIZATION} (UNCOVER) survey Data Release 4 \citep{2024ApJS..270....7W, 2024ApJ...976..101S, 2023MNRAS.523.4568F, 2024ApJS..270...12W} in the Abell 2744 field. We summarize the key properties of the galaxy \uncovgal in Fig. \ref{fig:spectrum}, including the multi-band JWST/NIRCam cutouts \citep{2024ApJ...976..101S, 2024ApJ...974...92B}, the observed 1D NIRSpec PRISM spectrum \citep{2025ApJ...982...51P}, and the population synthesis modeling results. The galaxy can be identified with {\tt msa\_id} = 3686 in the UNCOVER survey DR4 catalog. \uncovgal, also identified as $\texttt{Gz9p3}$, was first reported by \cite{boyett2024massive} from the GLASS-JWST survey, where NIRSpec/PRISM spectroscopy covered $\sim 1.1$-$4.5~\mu$m (with detector gaps). We further notice, at the time of writing the paper, that deeper spectroscopic observations of \uncovgal has been obtained by the \texttt{SPURS} team, and the galaxy is identified as \texttt{SPURS-A2744-7} \citep{chen2026spurs} in their work. In the present paper, we use the UNCOVER NIRSpec/PRISM observations presented by \cite{2025ApJ...982...51P}. The NIRSpec/PRISM Micro-Shutter Assembly (MSA) slit was positioned across the full extent of the galaxy and provides continuous wavelength coverage over $0.6$-$5.2~\mu$m and detect several additional emission lines, including H$\gamma$, and $[$O{\sc iii}$]\, (\lambda\lambda4960,5008)$.
The galaxy has a modest magnification ($\mu =1.6$, due to gravitational lensing). We show the NIRCam multi-band images \citep{2024ApJ...976..101S, 2024ApJ...974...92B} of the galaxy in Figure \ref{fig:spectrum} (panel 1A). The weak flux in the F115W filter, combined with clear detections in redder bands, is consistent with flux drop-out signatures of galaxies at $z>9$ due to Ly-$\alpha$ absorption \citep{atek2023revealing}. We perform the morphological measurement of the galaxy in the F277W filter by fitting a single S\'{e}rsic component. We find that the central light distribution is parametrized by axis ratio; $q=0.24$, S\'{e}rsic index; $n=1.4$, and an effective radius; $R_{e}=0.45$ kpc, indicating a compact disc morphology. The galaxy also exhibits a prominent, elongated tail extending up to $3.9$ kpc from its center, which could be a signature of an ongoing tidal interaction \citep{boyett2024massive}. We discuss the details of morphological fitting in Appendix \ref{secA3}.

\subsection{Spectral modeling} \label{subsec:spec-model}
We show the $1D$ JWST/NIRSpec PRISM spectrum+model of \texttt{UNCOVER 3686} in Fig. \ref{fig:spectrum} (panel 1B) from \cite{2025ApJ...982...51P}. The JWST/NIRSpec PRISM catalog provides the best-fit redshift to be $z_{\rm spec}=9.31^{+0.01}_{-0.01}$. In Fig. \ref{fig:spectrum} (panel 1B), we can see that \uncovgal unambiguously shows a strong Lyman break. The galaxy also exhibits prominent [\oiii] $\lambda\lambda 4959,5007$ emission lines indicating an actively star-forming environment, consistent with expectations for early galaxies during the epoch of reionization \citep{2014MNRAS.442..900N,suzuki2016iii,wold2025uncovering,laseter2024jades,2018ARA&A..56..435W}.
We measure the UV continuum slope to be $\beta=-1.93\pm0.06$ ($F_\lambda \propto \lambda^{\beta}$), which is redder than the median value ($\beta=-2.33$) estimated by \cite{2025arXiv250708245T} for a sample of 60 spectroscopically confirmed galaxies at $z>9$. This means that \uncovgal contains a moderate amount of dust, which is also evident from its dust attenuation ($A_V \approx 0.6$). However, such galaxies are not uncommon at these redshifts, as \cite{2025arXiv250708245T} also find 5 galaxies with slopes $\beta >-1.5$ at $z\sim 9$. More recent studies by \cite{algera2025first}, 
derive upper limits on dust mass $\rm <1.6\times10^{6}M_{\odot}$, suggesting that the \uncovgal is 
indeed deficient in its dust content. We find that \uncovgal has a high value of [\oiii]$\lambda 5007/H\beta=14.5$ and [\oiii]$\lambda\lambda$ 4959,5007/ [\oii]$\lambda 3727 = 11.9$, placing \texttt{UNCOVER 3686} in the AGN regime of the BPT diagram \citep{sanders2023excitation,fujimoto2024uncover}. However, \texttt{UNCOVER 3686} does not exhibit a broad H$\beta$ or [\oiii] component, thus ruling out the presence of an AGN \citep{bik2026spatially}. We infer the ionization parameter to be $\log U =-2.2$ using  [\neiii] $\lambda 3867$/[\oii]$\lambda 3727$ line ratio \citep{witstok2021assessing}. 

We infer the physical properties of \texttt{UNCOVER 3686} through a combined photometric and spectroscopic spectral energy distribution (SED) fitting of \emph{JWST} NIRCam+HST photometry together with the 1D NIRSpec PRISM spectrum using the \texttt{BAGPIPES} package \citep{carnall2018inferring}. We find that \uncovgal has stellar mass $\log_{10}(M_{\star}/M_{\odot}) = 9.55^{+0.08}_{-0.07}$, making the galaxy the most massive spectroscopically confirmed galaxy at $z>9$.  The stellar population is very young, with a mass-weighted age of $160^{+40}_{-30}$ Myr. The galaxy exhibits low stellar metallicity, $Z_{\star}/Z_{\odot} = 0.20^{+0.02}_{-0.02}$, and moderate dust attenuation, $A_V = 0.60^{+0.17}_{-0.15}$~mag, see Fig \ref{fig:spectrum} (panel C). The moderate dust attenuation is consistent with a slightly redder UV continuum slope. The inferred Lyman continuum escape fraction is $f_{\rm esc} = 0.50^{+0.04}_{-0.04}$, suggesting a significant leakage, which is estimated using the picket fence method \citep{2025arXiv250701096G} implemented in \texttt{BAGPIPES}. \texttt{UNCOVER 3686} is characterized by a very high average star formation rate of $29.8^{+4.7}_{-3.6}$ \mpy and a high specific star formation rate of $10^{-7.9}$ yr$^{-1}$. The high specific star formation rates indicate that the galaxy can add another $\sim 10^{9.5}M_{\odot}$ of stellar mass in the next $100$ Myr. Further details about the SED fitting are given in the Appendix \ref{secA1}. 

\begin{table}
\centering
\caption{Comparison of the observed properties of \uncovgal\ with the predictions of the feedback-free starburst (FFB) scenario from \citet{dekel2023efficient} and \citet{li2024feedback}.}
\label{tab:ffb_comparison}
\setlength{\tabcolsep}{2pt}
\begin{tabular}{lcc}
\hline\hline
Property & \uncovgal & FFB Prediction \\
\hline
Stellar mass $[\rm \log(M_\star/M_\odot)]$       & $9.55^{+0.08}_{-0.07}$   & $\sim 10$ \\
SFR [M$_\odot$\,yr$^{-1}$]                       & $29.8^{+4.7}_{-3.6}$     & $\sim65$  \\
Effective radius $[R_{\rm e}]$ [kpc]             & $0.45$                   & $\sim0.3$--$0.5$ \\
Metallicity $[Z_\star/Z_\odot]$                  & $0.20^{+0.02}_{-0.02}$   & $\lesssim0.1$ \\
Dust attenuation $[A_V]$                         & $0.60^{+0.17}_{-0.15}$   & $A_{\rm UV}\sim0.5$ \\
\hline\hline
\end{tabular}
\tablefoot{The FFB scenario predicts a threshold halo mass of $\log(M_{\rm h,FFB}/M_\odot)\simeq10.72$ at $z=9.3$, above which galaxies are expected to undergo feedback-free starbursts. The predicted quantities are taken from \citet{dekel2023efficient} and \citet{li2024feedback}. The galaxy has a stellar age equal to $1\rm 60^{+40}_{-30} Myr$ and the typical FFB phase lasts for $\rm \sim80$-$100 Myr$, so the stellar population is expected to be at least older than $\rm \sim80$-$100 Myr$.}
\end{table}

We summarize the physical properties of \uncovgal derived from spectrophotometric fitting and compare it with the quantitative predictions of the FFB scenario from  \cite{dekel2023efficient} and \cite{li2024feedback} in Table~\ref{tab:ffb_comparison}. The physical properties of \uncovgal are consistent with the FFB predictions across every observable property, such as effective radius, stellar age, metallicity, and dust attenuation. Having established that the physical properties of \uncovgal are consistent with the predictions of FFB, we now address the principal prediction of the FFB scenario: \emph{does \uncovgal\ attain the high baryon conversion efficiency expected for a feedback-free starburst galaxy?}

\section{Star formation efficiency} \label{sec:results}
The maximum amount of baryons that a dark matter halo can accommodate, $M_{b} = f_{b}M_{h}$, where $M_{h}$ is the mass of the dark matter halo and $f_{b}=\Omega_{b}/\Omega_{m} \approx 0.16$ is the cosmic baryon fraction. This consequently sets a ceiling on the stellar mass $M_\star(M_{h})\leq M_{b}(M_{h})$ that can form in the given halo. For a general baryon conversion efficiency, the relation becomes $M_{\star}=\varepsilon f_{b}  M _{h}$. To place \texttt{UNCOVER\,3686} in the context of the broader galaxy population at $z>9$, we construct a statistical reference sample at $9<z<10$ from the \texttt{ASTRODEEP-JWST} photometric catalog \citep{merlin2024astrodeep}, which spans six JWST extragalactic deep fields over $\simeq 0.2~\mathrm{deg}^2$. Of these six fields, only the \texttt{UNCOVER} pointing targets the lensing cluster Abell~2744; the remaining five are blank fields. The \texttt{UNCOVER} field covers $\sim$45~arcmin$^2$ in the image plane, corresponding to an effective source-plane area of $\sim$35~arcmin$^2$ \citep{atek2023revealing}. This source-plane area contributes less than 5\% of the total survey footprint of $\simeq 0.2$~deg$^2$ with a negligible effect on the population level abundance matching. In previous work, \citet{boyett2024massive} argued that \uncovgal\ is a rare object, with a selection volume bounded between the NIRSpec field of view ($\rm \sim 3 \times 3 arcmin{2}$) and the area covered by the ancillary HST/WFC3 imaging ($\rm \sim 13 arcmin^{2}$), and inferred qualitatively that such rare galaxies require unusually efficient star formation. Their analysis relied on the semi-empirical $M_{\rm UV}$-$M_{\rm h}$ framework of 
\citet{mason2015galaxy,mason2023brightest}, in which the star-formation efficiency, $\epsilon(M_{\rm h})$, is calibrated primarily as a function of halo mass using a reference sample at $\rm z\sim5$ and extrapolated to higher redshift. Our analysis addresses this directly by performing abundance matching at $\rm 9<z<10$ using an independently constructed reference sample spanning $\sim0.2\,\mathrm{deg}^{2}$, which is more than $\rm 50$ times the survey area used to bound the volume estimate of \citet{boyett2024massive}, with per-galaxy completeness and cosmic variance corrections, yielding the first quantitative measurement of $\epsilon$ for a spectroscopically confirmed galaxy at $z>9$. From an initial selection of 1757 galaxies with photometric redshifts from \texttt{PHOTZ} \citep{fontana2000photometric}, we derive stellar masses by fitting NIRCam photometry with \texttt{BAGPIPES} at fixed photometric redshifts, and apply the following quality cuts: $S/N<2$ for non-detection in F090W and F115W to confirm the Lyman break,  $S/N>4.5$ in at least two red NIRCam filters to ensure robust detection, $\chi^{2}_{\rm SED}<20$ to retain well fitted SEDs \citep{lee2024morphology}, and independent redshift agreement within  $9<z<10$ across four photometric redshift codes (\texttt{PHOTZ}, \texttt{EAzY\,v1.3}, \texttt{EAzY Ly$\alpha$ Reduced}, and \texttt{EAzY Ly$\alpha$}; \cite{brammer2008eazy,larson2023spectral}). This yields our primary \texttt{Robust-photo-z} sample of 142 galaxies, which includes 31 spectroscopically confirmed galaxies. The above strict selection criteria remove contaminants but risks discarding genuine high-z galaxies with marginally inconsistent photo-z estimates, whereas a relaxed selection retains completeness but may admit interlopers that inflate the high-mass end and bias halo mass assignments. To bracket this uncertainty, we construct another set of \texttt{Relaxed-photo-z} samples, constructed by retaining the same Lyman-break and $S/N$ criteria as the \texttt{Robust-photo-z} sample but dropping the multi-code redshift agreement requirement. The \texttt{Relaxed-photo-z} sample contains 721 galaxies; a fraction of them may be contaminants.

\begin{figure*}
\includegraphics[width=0.98\textwidth, height=10.0cm]{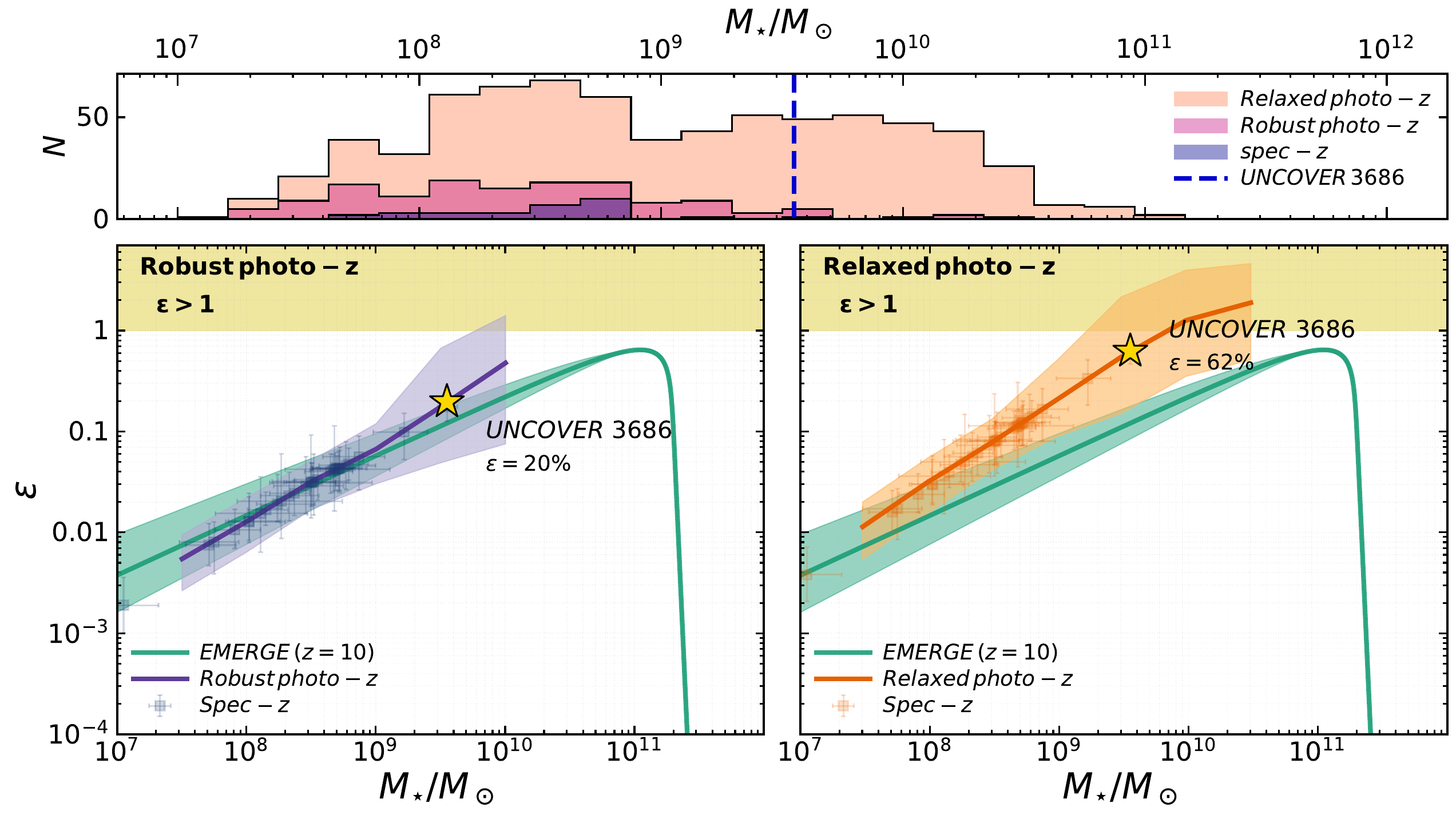}
\caption{
Star-formation efficiency, $\varepsilon = M_\star/(f_{\rm b}M_{\rm h})$, as a function of stellar mass derived from abundance matching for galaxies at $9<z<10$. The upper panel shows the stellar mass distributions of the \texttt{Robust-photo-z} sample (magenta), \texttt{Relaxed-photo-z} sample (orange), and spectroscopically confirmed galaxies (violet); the dashed blue line marks UNCOVER-3686 ($\log M_\star/M_\odot = 9.55$). The lower panels show efficiencies derived using the \texttt{Robust-photo-z} sample (left) and \texttt{Relaxed-photo-z} sample (right). The solid magenta and orange curves depict the running median (50th percentile) of $\varepsilon$ in bins of stellar mass, and the shaded region indicates the 16th–84th percentile range. Square markers indicate spectroscopically confirmed galaxies. The green curves and corresponding shaded regions show the \textsc{emerge} prediction at $z=10$ \citep{2018MNRAS.477.1822M}. The yellow stars mark the location of \uncovgal.}
\label{fig:efficiency}
\end{figure*}

We perform abundance matching using the Sheth-Tormen halo mass function \citep{sheth1999large} at $z=9.3$ with \textit{Planck} 2018 cosmological parameters \citep{aghanim2020planck}, applying per-galaxy completeness corrections from the \texttt{ASTRODEEP-JWST} detection curves \citep{merlin2024astrodeep} and cosmic variance estimates at the matched halo mass from \texttt{galcv} \citep{trapp2020flexible}.
We discuss the details of abundance matching in Appendix \ref{secA2}. We show the inferred efficiencies as a function of stellar mass for our primary \texttt{Robust-photo-z} sample in Fig. \ref{fig:efficiency} (left-panel). This method assigns \texttt{UNCOVER\,3686} a halo mass of $\log(M_h/M_\odot)=11$ and a star formation efficiency of $\varepsilon = 0.20^{+0.05}_{-0.06}$, a factor of three to four above \textsc{emerge} predictions \citep{2018MNRAS.477.1822M}. This value ($\rm \epsilon=0.22^{+0.06}_{-0.06}$) is independently confirmed by treating \texttt{UNCOVER\,3686} as the most massive spectroscopically confirmed galaxy in the \texttt{UNCOVER} pointing, using the delensed source plane volume $\sim$35 arcmin$^{2}$ \citep{atek2023revealing}. The feedback-free (FFB) scenario predicts a threshold halo mass $M_{h,\rm FFB} = 10^{10.8} \left[(1+z)/10\right]^{-6.2} M_\odot$, which equals $\log(M_h/M_\odot)=10.72$ at $z=9.3$. Therefore, \uncovgal exceeds the FFB threshold halo mass. For the \texttt{Relaxed-photo-z} sample, the abundance matching assigns a halo mass of $\log(M_h/M_\odot)=10.56$ to \texttt{UNCOVER\,3686}, which translates into $\varepsilon \approx 62\%$ (shown in the right panel of Fig. \ref{fig:efficiency}), even closer to the FFB predictions.

We further estimate the physical conditions of the \uncovgal. We compute the mean gas density by assuming that the observed stellar mass was initially in gaseous form within the effective radius, $R_e = 0.45$ kpc: $n_{\rm gas} \sim (3M_\star)/(4\pi R_e^3) \simeq 3.3 \times 10^2$ \pcc. Adopting a cloud contrast ratio of $\sim 10$ \citep{dekel2023efficient}, the characteristic cloud density reaches $n_{\rm gas} \sim 3\times 10^3$ \pcc, the threshold above which stellar winds and radiation-driven feedback are expected to be suppressed \citep{grudic2022dynamics,menon2022infrared}. Therefore, the size, gas density, the metallicity ($\approx 0.2\: Z_\odot$), and the young age ($\sim 160$ Myr) of the galaxy indicate the possibility that \uncovgal may be an excellent candidate for an FFB galaxy. Although the estimated range of SFE, $\varepsilon = 20\%-60\%$, is less than the near maximal efficiency ($\varepsilon=1$) in the FFB regime, we should note that the maximal efficiency in the FFB regime is expected only for brief periods of intense star formation in dense clouds and may not hold for the entire age of the galaxy. In practice, the average SFE in FFB galaxies would be determined by the associated duty cycle of such starburst periods, and hence a global $\varepsilon \approx 20\%-60\%$ would mean a characteristic starburst peak of $2-6$ Myr against a $10$ Myr star formation cycle \citep{li2024feedback}.  Having said that, we also acknowledge the fact that the higher end of the SFE may contain possible interlopers, as has been reported at high-$z$ photometric samples \citep{kocevski2023hidden, harikane2023comprehensive, arrabal2023confirmation, 2023ApJ...951L..22A}. In such a case, the global SFE estimate of \uncovgal would collapse to $\varepsilon \approx 20\%$, but would still remain within the FFB regime. Interestingly, recent deep spectroscopic observations of \uncovgal\ by the \texttt{SPURS} program \citep{chen2026spurs} detect outflows at $v \simeq -161$~km~s$^{-1}$ using interstellar absorption lines, suggesting that stellar feedback is not entirely absent in \uncovgal, suggesting that stellar feedback may have begun to set in and that \uncovgal\ may be observed at the edge of feedback-free star formation efficiency.

\section{Conclusions}\label{sec:conclusions}
We perform the first direct observational test of whether a galaxy at cosmic dawn that satisfies the physical conditions predicted for feedback-free star formation actually reaches the baryon conversion 
efficiency predicted by the FFB model \citep{dekel2023efficient, li2024feedback}. Using \uncovgal\ ($z = 9.31$) as an anchor against a reference sample of 142 galaxies containing 31 spectroscopically confirmed galaxies between $9 < z < 10$ spanning $0.2\mathrm{deg}^2$, we answer this question as follows.
\begin{enumerate}

\item The physical properties of \uncovgal compact morphology ($R_e = 0.45$~kpc), young stellar age ($\sim 160$~Myr), low metallicity  ($Z_\star/Z_\odot \approx 0.2$), and characteristic 
cloud density ($n_{\rm gas} \sim 3 \times 10^3$~cm$^{-3}$), are consistent with the predictions of the feedback-free starburst scenario \citep{dekel2023efficient, li2024feedback}, as summarized in Table~\ref{tab:ffb_comparison}.

\item Abundance matching assigns \uncovgal\ a halo mass of $\log(M_h/M_\odot) = 11$, exceeding the FFB threshold of $\log(M_{h,\rm FFB}/M_\odot) = 10.72$  at $z = 9.3$, and a star formation efficiency of $\varepsilon = 0.20^{+0.05}_{-0.06}$, a factor of three to four above \textsc{emerge} predictions \citep{2018MNRAS.477.1822M} and consistent with duty-cycle-averaged global efficiency  FFB predictions \citep{li2024feedback}. This is independently confirmed by treating \uncovgal\ as the most massive spectroscopically confirmed galaxy in the \texttt{UNCOVER} pointing using the delensed source-plane volume.

\item Recent deep spectroscopic observations by the \texttt{SPURS} program \citep{chen2026spurs}  detect interstellar outflows at $v \simeq -161$~km~s$^{-1}$, suggesting that 
stellar feedback may have begun to set in. This is consistent with the picture that \uncovgal\ is observed at the edge of feedback-free star formation efficiency, where the onset 
of feedback is beginning to suppress baryon conversion below the theoretical FFB maximum.

\item Whether the higher efficiency ($\approx 60\%$) implied by the \texttt{Relaxed-photo-z} sample is physical or photometric in origin remains an open question. If spectroscopically confirmed, they would establish a genuine near-maximal efficiency for a galaxy population at $z \sim 10$, while refutation would instead consolidate the SFE to $\approx 20\%$. Either outcome carries direct implications for theoretical models of feedback and baryonic cycle at the cosmic dawn. We further underscore the necessity of spectroscopic follow-up of the photometrically detected early massive galaxies in establishing robust star formation efficiency measurements.
\end{enumerate}

\begin{acknowledgements}
K. Aditya acknowledges support from FONDECYT Postdoctoral Fellowship (Project No. 3261078) and $\rm ANID-MILENIO-NCN2024\_112$. We dedicate this paper to the memory of Avishai Dekel, whose pioneering work on galaxy formation in the early Universe has profoundly shaped the ideas explored in this study. We also thank Bingjie Wang, Emiliano Merlin, Paola Santini, Mengtao Tang, and Jasjeet Singh Bagla for useful discussions. We thank Adam Carnall for his help with BAGPIPES. We thank the UNCOVER and ASTRODEEP teams for making their data publicly available, without which this study would not have been possible.
\end{acknowledgements}

\bibliographystyle{aasjournal}
\bibliography{2366_d2.bib}{}

@ARTICLE{2023NatAs...7..622C,
       author = {{Curtis-Lake}, Emma and {Carniani}, Stefano and {Cameron}, Alex and {Charlot}, Stephane and {Jakobsen}, Peter and {Maiolino}, Roberto and {Bunker}, Andrew and {Witstok}, Joris and {Smit}, Renske and {Chevallard}, Jacopo and {Willott}, Chris and {Ferruit}, Pierre and {Arribas}, Santiago and {Bonaventura}, Nina and {Curti}, Mirko and {D'Eugenio}, Francesco and {Franx}, Marijn and {Giardino}, Giovanna and {Looser}, Tobias J. and {L{\"u}tzgendorf}, Nora and {Maseda}, Michael V. and {Rawle}, Tim and {Rix}, Hans-Walter and {Rodr{\'\i}guez del Pino}, Bruno and {{\"U}bler}, Hannah and {Sirianni}, Marco and {Dressler}, Alan and {Egami}, Eiichi and {Eisenstein}, Daniel J. and {Endsley}, Ryan and {Hainline}, Kevin and {Hausen}, Ryan and {Johnson}, Benjamin D. and {Rieke}, Marcia and {Robertson}, Brant and {Shivaei}, Irene and {Stark}, Daniel P. and {Tacchella}, Sandro and {Williams}, Christina C. and {Willmer}, Christopher N.~A. and {Bhatawdekar}, Rachana and {Bowler}, Rebecca and {Boyett}, Kristan and {Chen}, Zuyi and {de Graaff}, Anna and {Helton}, Jakob M. and {Hviding}, Raphael E. and {Jones}, Gareth C. and {Kumari}, Nimisha and {Lyu}, Jianwei and {Nelson}, Erica and {Perna}, Michele and {Sandles}, Lester and {Saxena}, Aayush and {Suess}, Katherine A. and {Sun}, Fengwu and {Topping}, Michael W. and {Wallace}, Imaan E.~B. and {Whitler}, Lily},
        title = "{Spectroscopic confirmation of four metal-poor galaxies at z = 10.3-13.2}",
      journal = {Nature Astronomy},
         year = 2023,
        month = may,
       volume = {7},
        pages = {622-632},
          doi = {10.1038/s41550-023-01918-w},
archivePrefix = {arXiv},
       eprint = {2212.04568},
 primaryClass = {astro-ph.GA},
       adsurl = {https://ui.adsabs.harvard.edu/abs/2023NatAs...7..622C}}

@ARTICLE{2023ApJ...951L..22A,
       author = {{Arrabal Haro}, Pablo and {Dickinson}, Mark and {Finkelstein}, Steven L. and {Fujimoto}, Seiji and {Fern{\'a}ndez}, Vital and {Kartaltepe}, Jeyhan S. and {Jung}, Intae and {Cole}, Justin W. and {Burgarella}, Denis and {Chworowsky}, Katherine and {Hutchison}, Taylor A. and {Morales}, Alexa M. and {Papovich}, Casey and {Simons}, Raymond C. and {Amor{\'\i}n}, Ricardo O. and {Backhaus}, Bren E. and {Bagley}, Micaela B. and {Bisigello}, Laura and {Calabr{\`o}}, Antonello and {Castellano}, Marco and {Cleri}, Nikko J. and {Dav{\'e}}, Romeel and {Dekel}, Avishai and {Ferguson}, Henry C. and {Fontana}, Adriano and {Gawiser}, Eric and {Giavalisco}, Mauro and {Harish}, Santosh and {Hathi}, Nimish P. and {Hirschmann}, Michaela and {Holwerda}, Benne W. and {Huertas-Company}, Marc and {Koekemoer}, Anton M. and {Larson}, Rebecca L. and {Lucas}, Ray A. and {Mobasher}, Bahram and {P{\'e}rez-Gonz{\'a}lez}, Pablo G. and {Pirzkal}, Nor and {Rose}, Caitlin and {Santini}, Paola and {Trump}, Jonathan R. and {de la Vega}, Alexander and {Wang}, Xin and {Weiner}, Benjamin J. and {Wilkins}, Stephen M. and {Yang}, Guang and {Yung}, L.~Y. Aaron and {Zavala}, Jorge A.},
        title = "{Spectroscopic Confirmation of CEERS NIRCam-selected Galaxies at z = 8-10}",
      journal = {\apjl},
         year = 2023,
        month = jul,
       volume = {951},
       number = {1},
          eid = {L22},
        pages = {L22},
          doi = {10.3847/2041-8213/acdd54},
archivePrefix = {arXiv},
       eprint = {2304.05378},
 primaryClass = {astro-ph.GA},
       adsurl = {https://ui.adsabs.harvard.edu/abs/2023ApJ...951L..22A}
}

@ARTICLE{2023MNRAS.523.1009B,
       author = {{Bouwens}, Rychard and {Illingworth}, Garth and {Oesch}, Pascal and {Stefanon}, Mauro and {Naidu}, Rohan and {van Leeuwen}, Ivana and {Magee}, Dan},
        title = "{UV luminosity density results at z > 8 from the first JWST/NIRCam fields: limitations of early data sets and the need for spectroscopy}",
      journal = {\mnras},
         year = 2023,
        month = jul,
       volume = {523},
       number = {1},
        pages = {1009-1035},
          doi = {10.1093/mnras/stad1014},
archivePrefix = {arXiv},
       eprint = {2212.06683},
 primaryClass = {astro-ph.CO},
       adsurl = {https://ui.adsabs.harvard.edu/abs/2023MNRAS.523.1009B}
}

@ARTICLE{2024ApJ...973....8H,
       author = {{Hsiao}, Tiger Yu-Yang and {Abdurro'uf} and {Coe}, Dan and {Larson}, Rebecca L. and {Jung}, Intae and {Mingozzi}, Matilde and {Dayal}, Pratika and {Kumari}, Nimisha and {Kokorev}, Vasily and {Vikaeus}, Anton and {Brammer}, Gabriel and {Furtak}, Lukas J. and {Adamo}, Angela and {Andrade-Santos}, Felipe and {Antwi-Danso}, Jacqueline and {Brada{\v{c}}}, Maru{\v{s}}a and {Bradley}, Larry D. and {Broadhurst}, Tom and {Carnall}, Adam C. and {Conselice}, Christopher J. and {Diego}, Jose M. and {Donahue}, Megan and {Eldridge}, Jan J. and {Fujimoto}, Seiji and {Henry}, Alaina and {Hernandez}, Svea and {Hutchison}, Taylor A. and {James}, Bethan L. and {Norman}, Colin and {Park}, Hyunbae and {Pirzkal}, Norbert and {Postman}, Marc and {Ricotti}, Massimo and {Rigby}, Jane R. and {Vanzella}, Eros and {Welch}, Brian and {Wilkins}, Stephen M. and {Windhorst}, Rogier A. and {Xu}, Xinfeng and {Zackrisson}, Erik and {Zitrin}, Adi},
        title = "{JWST NIRSpec Spectroscopy of the Triply Lensed z = 10.17 Galaxy MACS0647{\textendash}JD}",
      journal = {\apj},
         year = 2024,
        month = sep,
       volume = {973},
       number = {1},
          eid = {8},
        pages = {8},
          doi = {10.3847/1538-4357/ad5da8},
archivePrefix = {arXiv},
       eprint = {2305.03042},
 primaryClass = {astro-ph.GA},
       adsurl = {https://ui.adsabs.harvard.edu/abs/2024ApJ...973....8H}
}

@ARTICLE{2023Natur.618..480R,
       author = {{Roberts-Borsani}, Guido and {Treu}, Tommaso and {Chen}, Wenlei and {Morishita}, Takahiro and {Vanzella}, Eros and {Zitrin}, Adi and {Bergamini}, Pietro and {Castellano}, Marco and {Fontana}, Adriano and {Glazebrook}, Karl and {Grillo}, Claudio and {Kelly}, Patrick L. and {Merlin}, Emiliano and {Nanayakkara}, Themiya and {Paris}, Diego and {Rosati}, Piero and {Yang}, Lilan and {Acebron}, Ana and {Bonchi}, Andrea and {Boyett}, Kit and {Brada{\v{c}}}, Maru{\v{s}}a and {Brammer}, Gabriel and {Broadhurst}, Tom and {Calabr{\'o}}, Antonello and {Diego}, Jose M. and {Dressler}, Alan and {Furtak}, Lukas J. and {Filippenko}, Alexei V. and {Henry}, Alaina and {Koekemoer}, Anton M. and {Leethochawalit}, Nicha and {Malkan}, Matthew A. and {Mason}, Charlotte and {Mercurio}, Amata and {Metha}, Benjamin and {Pentericci}, Laura and {Pierel}, Justin and {Rieck}, Steven and {Roy}, Namrata and {Santini}, Paola and {Strait}, Victoria and {Strausbaugh}, Robert and {Trenti}, Michele and {Vulcani}, Benedetta and {Wang}, Lifan and {Wang}, Xin and {Windhorst}, Rogier A.},
        title = "{The nature of an ultra-faint galaxy in the cosmic dark ages seen with JWST}",
      journal = {\nat},
         year = 2023,
        month = jun,
       volume = {618},
       number = {7965},
        pages = {480-483},
          doi = {10.1038/s41586-023-05994-w},
archivePrefix = {arXiv},
       eprint = {2210.15639},
 primaryClass = {astro-ph.GA},
       adsurl = {https://ui.adsabs.harvard.edu/abs/2023Natur.618..480R}
}

@ARTICLE{2023Sci...380..416W,
       author = {{Williams}, Hayley and {Kelly}, Patrick L. and {Chen}, Wenlei and {Brammer}, Gabriel and {Zitrin}, Adi and {Treu}, Tommaso and {Scarlata}, Claudia and {Koekemoer}, Anton M. and {Oguri}, Masamune and {Lin}, Yu-Heng and {Diego}, Jose M. and {Nonino}, Mario and {Hjorth}, Jens and {Langeroodi}, Danial and {Broadhurst}, Tom and {Rogers}, Noah and {Perez-Fournon}, Ismael and {Foley}, Ryan J. and {Jha}, Saurabh and {Filippenko}, Alexei V. and {Strolger}, Lou and {Pierel}, Justin and {Poidevin}, Frederick and {Yang}, Lilan},
        title = "{A magnified compact galaxy at redshift 9.51 with strong nebular emission lines}",
      journal = {Science},
         year = 2023,
        month = apr,
       volume = {380},
       number = {6643},
        pages = {416-420},
          doi = {10.1126/science.adf5307},
archivePrefix = {arXiv},
       eprint = {2210.15699},
 primaryClass = {astro-ph.GA},
       adsurl = {https://ui.adsabs.harvard.edu/abs/2023Sci...380..416W}
}

@ARTICLE{2023A&A...677A.115C,
       author = {{Cameron}, Alex J. and {Saxena}, Aayush and {Bunker}, Andrew J. and {D'Eugenio}, Francesco and {Carniani}, Stefano and {Maiolino}, Roberto and {Curtis-Lake}, Emma and {Ferruit}, Pierre and {Jakobsen}, Peter and {Arribas}, Santiago and {Bonaventura}, Nina and {Charlot}, Stephane and {Chevallard}, Jacopo and {Curti}, Mirko and {Looser}, Tobias J. and {Maseda}, Michael V. and {Rawle}, Tim and {Rodr{\'\i}guez Del Pino}, Bruno and {Smit}, Renske and {{\"U}bler}, Hannah and {Willott}, Chris and {Witstok}, Joris and {Egami}, Eiichi and {Eisenstein}, Daniel J. and {Johnson}, Benjamin D. and {Hainline}, Kevin and {Rieke}, Marcia and {Robertson}, Brant E. and {Stark}, Daniel P. and {Tacchella}, Sandro and {Williams}, Christina C. and {Willmer}, Christopher N.~A. and {Bhatawdekar}, Rachana and {Bowler}, Rebecca and {Boyett}, Kristan and {Circosta}, Chiara and {Helton}, Jakob M. and {Jones}, Gareth C. and {Kumari}, Nimisha and {Ji}, Zhiyuan and {Nelson}, Erica and {Parlanti}, Eleonora and {Sandles}, Lester and {Scholtz}, Jan and {Sun}, Fengwu},
        title = "{JADES: Probing interstellar medium conditions at z {\ensuremath{\sim}} 5.5-9.5 with ultra-deep JWST/NIRSpec spectroscopy}",
      journal = {\aap},
         year = 2023,
        month = sep,
       volume = {677},
          eid = {A115},
        pages = {A115},
          doi = {10.1051/0004-6361/202346107},
archivePrefix = {arXiv},
       eprint = {2302.04298},
 primaryClass = {astro-ph.GA},
       adsurl = {https://ui.adsabs.harvard.edu/abs/2023A&A...677A.115C}
}

@ARTICLE{2024NatAs...8..657B,
       author = {{Boyett}, Kristan and {Trenti}, Michele and {Leethochawalit}, Nicha and {Calabr{\'o}}, Antonello and {Metha}, Benjamin and {Roberts-Borsani}, Guido and {Dalmasso}, Nicol{\'o} and {Yang}, Lilan and {Santini}, Paola and {Treu}, Tommaso and {Jones}, Tucker and {Henry}, Alaina and {Mason}, Charlotte A. and {Morishita}, Takahiro and {Nanayakkara}, Themiya and {Roy}, Namrata and {Wang}, Xin and {Fontana}, Adriano and {Merlin}, Emiliano and {Castellano}, Marco and {Paris}, Diego and {Brada{\v{c}}}, Maru{\v{s}}a and {Malkan}, Matt and {Marchesini}, Danilo and {Mascia}, Sara and {Glazebrook}, Karl and {Pentericci}, Laura and {Vanzella}, Eros and {Vulcani}, Benedetta},
        title = "{A massive interacting galaxy 510 million years after the Big Bang}",
      journal = {Nature Astronomy},
         year = 2024,
        month = may,
       volume = {8},
        pages = {657-672},
          doi = {10.1038/s41550-024-02218-7},
archivePrefix = {arXiv},
       eprint = {2303.00306},
 primaryClass = {astro-ph.GA},
       adsurl = {https://ui.adsabs.harvard.edu/abs/2024NatAs...8..657B}
}

@ARTICLE{2018Natur.557..392H,
       author = {{Hashimoto}, Takuya and {Laporte}, Nicolas and {Mawatari}, Ken and {Ellis}, Richard S. and {Inoue}, Akio K. and {Zackrisson}, Erik and {Roberts-Borsani}, Guido and {Zheng}, Wei and {Tamura}, Yoichi and {Bauer}, Franz E. and {Fletcher}, Thomas and {Harikane}, Yuichi and {Hatsukade}, Bunyo and {Hayatsu}, Natsuki H. and {Matsuda}, Yuichi and {Matsuo}, Hiroshi and {Okamoto}, Takashi and {Ouchi}, Masami and {Pell{\'o}}, Roser and {Rydberg}, Claes-Erik and {Shimizu}, Ikkoh and {Taniguchi}, Yoshiaki and {Umehata}, Hideki and {Yoshida}, Naoki},
        title = "{The onset of star formation 250 million years after the Big Bang}",
      journal = {\nat},
         year = 2018,
        month = may,
       volume = {557},
       number = {7705},
        pages = {392-395},
          doi = {10.1038/s41586-018-0117-z},
archivePrefix = {arXiv},
       eprint = {1805.05966},
 primaryClass = {astro-ph.GA},
       adsurl = {https://ui.adsabs.harvard.edu/abs/2018Natur.557..392H}
}

@ARTICLE{2023MNRAS.526.1657T,
       author = {{Tang}, Mengtao and {Stark}, Daniel P. and {Chen}, Zuyi and {Mason}, Charlotte and {Topping}, Michael and {Endsley}, Ryan and {Senchyna}, Peter and {Plat}, Ad{\`e}le and {Lu}, Ting-Yi and {Whitler}, Lily and {Robertson}, Brant and {Charlot}, St{\'e}phane},
        title = "{JWST/NIRSpec spectroscopy of z = 7-9 star-forming galaxies with CEERS: new insight into bright Ly{\ensuremath{\alpha}} emitters in ionized bubbles}",
      journal = {\mnras},
         year = 2023,
        month = dec,
       volume = {526},
       number = {2},
        pages = {1657-1686},
          doi = {10.1093/mnras/stad2763},
archivePrefix = {arXiv},
       eprint = {2301.07072},
 primaryClass = {astro-ph.GA},
       adsurl = {https://ui.adsabs.harvard.edu/abs/2023MNRAS.526.1657T}
}

@ARTICLE{2023ApJ...955..130F,
       author = {{Fujimoto}, Seiji and {Finkelstein}, Steven L. and {Burgarella}, Denis and {Carilli}, Chris L. and {Buat}, V{\'e}ronique and {Casey}, Caitlin M. and {Ciesla}, Laure and {Tacchella}, Sandro and {Zavala}, Jorge A. and {Brammer}, Gabriel and {Fudamoto}, Yoshinobu and {Ouchi}, Masami and {Valentino}, Francesco and {Cooper}, M.~C. and {Dickinson}, Mark and {Franco}, Maximilien and {Giavalisco}, Mauro and {Hutchison}, Taylor A. and {Kartaltepe}, Jeyhan S. and {Koekemoer}, Anton M. and {Kojima}, Takashi and {Larson}, Rebecca L. and {Murphy}, E.~J. and {Papovich}, Casey and {P{\'e}rez-Gonz{\'a}lez}, Pablo G. and {Somerville}, Rachel S. and {Yoon}, Ilsang and {Wilkins}, Stephen M. and {Akins}, Hollis and {Amor{\'\i}n}, Ricardo O. and {Arrabal Haro}, Pablo and {Bagley}, Micaela B. and {Chworowsky}, Katherine and {Cleri}, Nikko J. and {Cooper}, Olivia R. and {Costantin}, Luca and {Daddi}, Emanuele and {Ferguson}, Henry C. and {Grogin}, Norman A. and {Jim{\'e}nez-Andrade}, E.~F. and {Juneau}, St{\'e}phanie and {Kirkpatrick}, Allison and {Kocevski}, Dale D. and {Le Bail}, Aur{\'e}lien and {Long}, Arianna and {Lucas}, Ray A. and {Magnelli}, Benjamin and {McKinney}, Jed and {Rose}, Caitlin and {Seill{\'e}}, Lise-Marie and {Simons}, Raymond C. and {Weiner}, Benjamin J. and {Yung}, L.~Y. Aaron},
        title = "{ALMA FIR View of Ultra-high-redshift Galaxy Candidates at z {\ensuremath{\sim}} 11-17: Blue Monsters or Low-z Red Interlopers?}",
      journal = {\apj},
         year = 2023,
        month = oct,
       volume = {955},
       number = {2},
          eid = {130},
        pages = {130},
          doi = {10.3847/1538-4357/aceb67},
archivePrefix = {arXiv},
       eprint = {2211.03896},
 primaryClass = {astro-ph.GA},
       adsurl = {https://ui.adsabs.harvard.edu/abs/2023ApJ...955..130F}
}

@ARTICLE{2023ApJS..269...33N,
       author = {{Nakajima}, Kimihiko and {Ouchi}, Masami and {Isobe}, Yuki and {Harikane}, Yuichi and {Zhang}, Yechi and {Ono}, Yoshiaki and {Umeda}, Hiroya and {Oguri}, Masamune},
        title = "{JWST Census for the Mass-Metallicity Star Formation Relations at z = 4-10 with Self-consistent Flux Calibration and Proper Metallicity Calibrators}",
      journal = {\apjs},
         year = 2023,
        month = dec,
       volume = {269},
       number = {2},
          eid = {33},
        pages = {33},
          doi = {10.3847/1538-4365/acd556},
archivePrefix = {arXiv},
       eprint = {2301.12825},
 primaryClass = {astro-ph.GA},
       adsurl = {https://ui.adsabs.harvard.edu/abs/2023ApJS..269...33N}
}

@ARTICLE{2015ApJ...810L..12Z,
       author = {{Zitrin}, Adi and {Labb{\'e}}, Ivo and {Belli}, Sirio and {Bouwens}, Rychard and {Ellis}, Richard S. and {Roberts-Borsani}, Guido and {Stark}, Daniel P. and {Oesch}, Pascal A. and {Smit}, Renske},
        title = "{Lyman{\ensuremath{\alpha}} Emission from a Luminous z = 8.68 Galaxy: Implications for Galaxies as Tracers of Cosmic Reionization}",
      journal = {\apjl},
         year = 2015,
        month = sep,
       volume = {810},
       number = {1},
          eid = {L12},
        pages = {L12},
          doi = {10.1088/2041-8205/810/1/L12},
archivePrefix = {arXiv},
       eprint = {1507.02679},
 primaryClass = {astro-ph.GA},
       adsurl = {https://ui.adsabs.harvard.edu/abs/2015ApJ...810L..12Z}
}

@ARTICLE{2024ApJ...962...24S,
       author = {{Sanders}, Ryan L. and {Shapley}, Alice E. and {Topping}, Michael W. and {Reddy}, Naveen A. and {Brammer}, Gabriel B.},
        title = "{Direct T $_{e}$-based Metallicities of z = 2{\textendash}9 Galaxies with JWST/NIRSpec: Empirical Metallicity Calibrations Applicable from Reionization to Cosmic Noon}",
      journal = {\apj},
         year = 2024,
        month = feb,
       volume = {962},
       number = {1},
          eid = {24},
        pages = {24},
          doi = {10.3847/1538-4357/ad15fc},
archivePrefix = {arXiv},
       eprint = {2303.08149},
 primaryClass = {astro-ph.GA},
       adsurl = {https://ui.adsabs.harvard.edu/abs/2024ApJ...962...24S}
}

@article{naidu2022two,
  title={Two remarkably luminous galaxy candidates at z= 10--12 revealed by JWST},
  author={Naidu, Rohan P and Oesch, Pascal A and van Dokkum, Pieter and Nelson, Erica J and Suess, Katherine A and Brammer, Gabriel and Whitaker, Katherine E and Illingworth, Garth and Bouwens, Rychard and Tacchella, Sandro and others},
  journal={The Astrophysical Journal Letters},
  volume={940},
  number={1},
  pages={L14},
  year={2022},
  publisher={IOP Publishing}
}

@article{glazebrook2024massive,
  title={A massive galaxy that formed its stars at z = 11},
  author={Glazebrook, Karl and Nanayakkara, Themiya and Schreiber, Corentin and Lagos, Claudia and Kawinwanichakij, Lalitwadee and Jacobs, Colin and Chittenden, Harry and Brammer, Gabriel and Kacprzak, Glenn G and Labbe, Ivo and others},
  journal={Nature},
  volume={628},
  number={8007},
  pages={277--281},
  year={2024},
  publisher={Nature Publishing Group UK London}
}

@article{atek2023revealing,
  title={Revealing galaxy candidates out to z~ 16 with JWST observations of the lensing cluster SMACS0723},
  author={Atek, Hakim and Shuntov, Marko and Furtak, Lukas J and Richard, Johan and Kneib, Jean-Paul and Mahler, Guillaume and Zitrin, Adi and McCracken, HJ and Charlot, St{\'e}phane and Chevallard, Jacopo and others},
  journal={Monthly Notices of the Royal Astronomical Society},
  volume={519},
  number={1},
  pages={1201--1220},
  year={2023},
  publisher={Oxford University Press}
}

@article{harikane2023comprehensive,
  title={A comprehensive study of galaxies at z~ 9--16 found in the early JWST data: ultraviolet luminosity functions and cosmic star formation history at the pre-reionization epoch},
  author={Harikane, Yuichi and Ouchi, Masami and Oguri, Masamune and Ono, Yoshiaki and Nakajima, Kimihiko and Isobe, Yuki and Umeda, Hiroya and Mawatari, Ken and Zhang, Yechi},
  journal={The Astrophysical Journal Supplement Series},
  volume={265},
  number={1},
  pages={5},
  year={2023},
  publisher={IOP Publishing}
}

@article{labbe2023population,
  title={A population of red candidate massive galaxies\~{} 600 Myr after the Big Bang},
  author={Labb{\'e}, Ivo and van Dokkum, Pieter and Nelson, Erica and Bezanson, Rachel and Suess, Katherine A and Leja, Joel and Brammer, Gabriel and Whitaker, Katherine and Mathews, Elijah and Stefanon, Mauro and others},
  journal={Nature},
  volume={616},
  number={7956},
  pages={266--269},
  year={2023},
  publisher={Nature Publishing Group UK London}
}

@article{xiao2024accelerated,
  title={Accelerated formation of ultra-massive galaxies in the first billion years},
  author={Xiao, Mengyuan and Oesch, Pascal A and Elbaz, David and Bing, Longji and Nelson, Erica J and Weibel, Andrea and Illingworth, Garth D and van Dokkum, Pieter and Naidu, Rohan P and Daddi, Emanuele and others},
  journal={Nature},
  volume={635},
  number={8038},
  pages={311--315},
  year={2024},
  publisher={Nature Publishing Group UK London}
}

@article{boylan2023stress,
  title={Stress testing $\Lambda$ CDM with high-redshift galaxy candidates},
  author={Boylan-Kolchin, Michael},
  journal={Nature Astronomy},
  volume={7},
  number={6},
  pages={731--735},
  year={2023},
  publisher={Nature Publishing Group UK London}
}

@article{kocevski2023hidden,
  title={Hidden little monsters: spectroscopic identification of low-mass, broad-line AGNs at z> 5 with CEERS},
  author={Kocevski, Dale D and Onoue, Masafusa and Inayoshi, Kohei and Trump, Jonathan R and Haro, Pablo Arrabal and Grazian, Andrea and Dickinson, Mark and Finkelstein, Steven L and Kartaltepe, Jeyhan S and Hirschmann, Michaela and others},
  journal={The Astrophysical Journal Letters},
  volume={954},
  number={1},
  pages={L4},
  year={2023},
  publisher={IOP Publishing}
}

@article{carnall2023massive,
  title={A massive quiescent galaxy at redshift 4.658},
  author={Carnall, Adam C and McLure, Ross J and Dunlop, James S and McLeod, Derek J and Wild, Vivienne and Cullen, Fergus and Magee, Dan and Begley, Ryan and Cimatti, Andrea and Donnan, Callum T and others},
  journal={Nature},
  volume={619},
  number={7971},
  pages={716--719},
  year={2023},
  publisher={Nature Publishing Group UK London}
}

@article{de2025efficient,
  title={Efficient formation of a massive quiescent galaxy at redshift 4.9},
  author={De Graaff, Anna and Setton, David J and Brammer, Gabriel and Cutler, Sam and Suess, Katherine A and Labb{\'e}, Ivo and Leja, Joel and Weibel, Andrea and Maseda, Michael V and Whitaker, Katherine E and others},
  journal={Nature Astronomy},
  volume={9},
  number={2},
  pages={280--292},
  year={2025},
  publisher={Nature Publishing Group UK London}
}

@article{arrabal2023confirmation,
  title={Confirmation and refutation of very luminous galaxies in the early Universe},
  author={Arrabal Haro, Pablo and Dickinson, Mark and Finkelstein, Steven L and Kartaltepe, Jeyhan S and Donnan, Callum T and Burgarella, Denis and Carnall, Adam C and Cullen, Fergus and Dunlop, James S and Fern{\'a}ndez, Vital and others},
  journal={Nature},
  volume={622},
  number={7984},
  pages={707--711},
  year={2023},
  publisher={Nature Publishing Group UK London}
}

@ARTICLE{2013MNRAS.428.3121M,
       author = {{Moster}, Benjamin P. and {Naab}, Thorsten and {White}, Simon D.~M.},
        title = "{Galactic star formation and accretion histories from matching galaxies to dark matter haloes}",
      journal = {\mnras},
         year = 2013,
        month = feb,
       volume = {428},
       number = {4},
        pages = {3121-3138},
          doi = {10.1093/mnras/sts261},
archivePrefix = {arXiv},
       eprint = {1205.5807},
 primaryClass = {astro-ph.CO},
       adsurl = {https://ui.adsabs.harvard.edu/abs/2013MNRAS.428.3121M}
}

@ARTICLE{2018MNRAS.477.1822M,
       author = {{Moster}, Benjamin P. and {Naab}, Thorsten and {White}, Simon D.~M.},
        title = "{EMERGE - an empirical model for the formation of galaxies since z {\ensuremath{\sim}} 10}",
      journal = {\mnras},
         year = 2018,
        month = jun,
       volume = {477},
       number = {2},
        pages = {1822-1852},
          doi = {10.1093/mnras/sty655},
archivePrefix = {arXiv},
       eprint = {1705.05373},
 primaryClass = {astro-ph.GA},
       adsurl = {https://ui.adsabs.harvard.edu/abs/2018MNRAS.477.1822M}
}

@ARTICLE{2018ApJ...868...92T,
       author = {{Tacchella}, Sandro and {Bose}, Sownak and {Conroy}, Charlie and {Eisenstein}, Daniel J. and {Johnson}, Benjamin D.},
        title = "{A Redshift-independent Efficiency Model: Star Formation and Stellar Masses in Dark Matter Halos at z {\ensuremath{\gtrsim}} 4}",
      journal = {\apj},
         year = 2018,
        month = dec,
       volume = {868},
       number = {2},
          eid = {92},
        pages = {92},
          doi = {10.3847/1538-4357/aae8e0},
archivePrefix = {arXiv},
       eprint = {1806.03299},
 primaryClass = {astro-ph.GA},
       adsurl = {https://ui.adsabs.harvard.edu/abs/2018ApJ...868...92T}
}

@ARTICLE{2018ARA&A..56..435W,
       author = {{Wechsler}, Risa H. and {Tinker}, Jeremy L.},
        title = "{The Connection Between Galaxies and Their Dark Matter Halos}",
      journal = {\araa},
         year = 2018,
        month = sep,
       volume = {56},
        pages = {435-487},
          doi = {10.1146/annurev-astro-081817-051756},
archivePrefix = {arXiv},
       eprint = {1804.03097},
 primaryClass = {astro-ph.GA},
       adsurl = {https://ui.adsabs.harvard.edu/abs/2018ARA&A..56..435W}
}

@ARTICLE{2024ApJS..270....7W,
       author = {{Weaver}, John R. and {Cutler}, Sam E. and {Pan}, Richard and {Whitaker}, Katherine E. and {Labb{\'e}}, Ivo and {Price}, Sedona H. and {Bezanson}, Rachel and {Brammer}, Gabriel and {Marchesini}, Danilo and {Leja}, Joel and {Wang}, Bingjie and {Furtak}, Lukas J. and {Zitrin}, Adi and {Atek}, Hakim and {Chemerynska}, Iryna and {Coe}, Dan and {Dayal}, Pratika and {van Dokkum}, Pieter and {Feldmann}, Robert and {F{\"o}rster Schreiber}, Natascha M. and {Franx}, Marijn and {Fujimoto}, Seiji and {Fudamoto}, Yoshinobu and {Glazebrook}, Karl and {de Graaff}, Anna and {Greene}, Jenny E. and {Juneau}, St{\'e}phanie and {Kassin}, Susan and {Kriek}, Mariska and {Khullar}, Gourav and {Maseda}, Michael V. and {Mowla}, Lamiya A. and {Muzzin}, Adam and {Nanayakkara}, Themiya and {Nelson}, Erica J. and {Oesch}, Pascal A. and {Pacifici}, Camilla and {Papovich}, Casey and {Setton}, David J. and {Shapley}, Alice E. and {Shipley}, Heath V. and {Smit}, Renske and {Stefanon}, Mauro and {Taylor}, Edward N. and {Weibel}, Andrea and {Williams}, Christina C.},
        title = "{The UNCOVER Survey: A First-look HST + JWST Catalog of 60,000 Galaxies near A2744 and beyond}",
      journal = {\apjs},
         year = 2024,
        month = jan,
       volume = {270},
       number = {1},
          eid = {7},
        pages = {7},
          doi = {10.3847/1538-4365/ad07e0},
archivePrefix = {arXiv},
       eprint = {2301.02671},
 primaryClass = {astro-ph.GA},
       adsurl = {https://ui.adsabs.harvard.edu/abs/2024ApJS..270....7W}
}

@ARTICLE{2024ApJ...976..101S,
       author = {{Suess}, Katherine A. and {Weaver}, John R. and {Price}, Sedona H. and {Pan}, Richard and {Wang}, Bingjie and {Bezanson}, Rachel and {Brammer}, Gabriel and {Cutler}, Sam E. and {Labb{\'e}}, Ivo and {Leja}, Joel and {Williams}, Christina C. and {Whitaker}, Katherine E. and {Atek}, Hakim and {Dayal}, Pratika and {de Graaff}, Anna and {Feldmann}, Robert and {Franx}, Marijn and {Fudamoto}, Yoshinobu and {Fujimoto}, Seiji and {Furtak}, Lukas J. and {Goulding}, Andy D. and {Greene}, Jenny E. and {Khullar}, Gourav and {Kokorev}, Vasily and {Kriek}, Mariska and {Lorenz}, Brian and {Marchesini}, Danilo and {Maseda}, Michael V. and {Matthee}, Jorryt and {Miller}, Tim B. and {Mitsuhashi}, Ikki and {Mowla}, Lamiya A. and {Muzzin}, Adam and {Naidu}, Rohan P. and {Nanayakkara}, Themiya and {Nelson}, Erica J. and {Oesch}, Pascal A. and {Setton}, David J. and {Shipley}, Heath and {Smit}, Renske and {Spilker}, Justin S. and {van Dokkum}, Pieter and {Zitrin}, Adi},
        title = "{Medium Bands, Mega Science: A JWST/NIRCam Medium-band Imaging Survey of A2744}",
      journal = {\apj},
         year = 2024,
        month = nov,
       volume = {976},
       number = {1},
          eid = {101},
        pages = {101},
          doi = {10.3847/1538-4357/ad75fe},
archivePrefix = {arXiv},
       eprint = {2404.13132},
 primaryClass = {astro-ph.GA},
       adsurl = {https://ui.adsabs.harvard.edu/abs/2024ApJ...976..101S}
}

@ARTICLE{2023MNRAS.523.4568F,
       author = {{Furtak}, Lukas J. and {Zitrin}, Adi and {Weaver}, John R. and {Atek}, Hakim and {Bezanson}, Rachel and {Labb{\'e}}, Ivo and {Whitaker}, Katherine E. and {Leja}, Joel and {Price}, Sedona H. and {Brammer}, Gabriel B. and {Wang}, Bingjie and {Marchesini}, Danilo and {Pan}, Richard and {Dayal}, Pratika and {van Dokkum}, Pieter and {Feldmann}, Robert and {Fujimoto}, Seiji and {Franx}, Marijn and {Khullar}, Gourav and {Nelson}, Erica J. and {Mowla}, Lamiya A.},
        title = "{UNCOVERing the extended strong lensing structures of Abell 2744 with the deepest JWST imaging}",
      journal = {\mnras},
         year = 2023,
        month = aug,
       volume = {523},
       number = {3},
        pages = {4568-4582},
          doi = {10.1093/mnras/stad1627},
archivePrefix = {arXiv},
       eprint = {2212.04381},
 primaryClass = {astro-ph.GA},
       adsurl = {https://ui.adsabs.harvard.edu/abs/2023MNRAS.523.4568F}
}

@ARTICLE{2024ApJS..270...12W,
       author = {{Wang}, Bingjie and {Leja}, Joel and {Labb{\'e}}, Ivo and {Bezanson}, Rachel and {Whitaker}, Katherine E. and {Brammer}, Gabriel and {Furtak}, Lukas J. and {Weaver}, John R. and {Price}, Sedona H. and {Zitrin}, Adi and {Atek}, Hakim and {Coe}, Dan and {Cutler}, Sam E. and {Dayal}, Pratika and {van Dokkum}, Pieter and {Feldmann}, Robert and {Marchesini}, Danilo and {Franx}, Marijn and {F{\"o}rster Schreiber}, Natascha and {Fujimoto}, Seiji and {Geha}, Marla and {Glazebrook}, Karl and {de Graaff}, Anna and {Greene}, Jenny E. and {Juneau}, St{\'e}phanie and {Kassin}, Susan and {Kriek}, Mariska and {Khullar}, Gourav and {Maseda}, Michael and {Mowla}, Lamiya A. and {Muzzin}, Adam and {Nanayakkara}, Themiya and {Nelson}, Erica J. and {Oesch}, Pascal A. and {Pacifici}, Camilla and {Pan}, Richard and {Papovich}, Casey and {Setton}, David J. and {Shapley}, Alice E. and {Smit}, Renske and {Stefanon}, Mauro and {Suess}, Katherine A. and {Taylor}, Edward N. and {Williams}, Christina C.},
        title = "{The UNCOVER Survey: A First-look HST+JWST Catalog of Galaxy Redshifts and Stellar Population Properties Spanning 0.2 {\ensuremath{\lesssim}} z {\ensuremath{\lesssim}} 15}",
      journal = {\apjs},
         year = 2024,
        month = jan,
       volume = {270},
       number = {1},
          eid = {12},
        pages = {12},
          doi = {10.3847/1538-4365/ad0846},
archivePrefix = {arXiv},
       eprint = {2310.01276},
 primaryClass = {astro-ph.GA},
       adsurl = {https://ui.adsabs.harvard.edu/abs/2024ApJS..270...12W}
}

@ARTICLE{2025ApJ...982...51P,
       author = {{Price}, Sedona H. and {Bezanson}, Rachel and {Labbe}, Ivo and {Furtak}, Lukas J. and {de Graaff}, Anna and {Greene}, Jenny E. and {Kokorev}, Vasily and {Setton}, David J. and {Suess}, Katherine A. and {Brammer}, Gabriel and {Cutler}, Sam E. and {Leja}, Joel and {Pan}, Richard and {Wang}, Bingjie and {Weaver}, John R. and {Whitaker}, Katherine E. and {Atek}, Hakim and {Burgasser}, Adam J. and {Chemerynska}, Iryna and {Dayal}, Pratika and {Feldmann}, Robert and {F{\"o}rster Schreiber}, Natascha M. and {Fudamoto}, Yoshinobu and {Fujimoto}, Seiji and {Glazebrook}, Karl and {Goulding}, Andy D. and {Khullar}, Gourav and {Kriek}, Mariska and {Marchesini}, Danilo and {Maseda}, Michael V. and {Miller}, Tim B. and {Muzzin}, Adam and {Nanayakkara}, Themiya and {Nelson}, Erica and {Oesch}, Pascal A. and {Shipley}, Heath and {Smit}, Renske and {Taylor}, Edward N. and {Dokkum}, Pieter van and {Williams}, Christina C. and {Zitrin}, Adi},
        title = "{The UNCOVER Survey: First Release of Ultradeep JWST/NIRSpec PRISM Spectra for {\ensuremath{\sim}}700 Galaxies from z {\ensuremath{\sim}} 0.3{\textendash}13 in A2744}",
      journal = {\apj},
         year = 2025,
        month = mar,
       volume = {982},
       number = {1},
          eid = {51},
        pages = {51},
          doi = {10.3847/1538-4357/adaec1},
archivePrefix = {arXiv},
       eprint = {2408.03920},
 primaryClass = {astro-ph.GA},
       adsurl = {https://ui.adsabs.harvard.edu/abs/2025ApJ...982...51P}
}

@ARTICLE{2024ApJ...974...92B,
       author = {{Bezanson}, Rachel and {Labbe}, Ivo and {Whitaker}, Katherine E. and {Leja}, Joel and {Price}, Sedona H. and {Franx}, Marijn and {Brammer}, Gabriel and {Marchesini}, Danilo and {Zitrin}, Adi and {Wang}, Bingjie and {Weaver}, John R. and {Furtak}, Lukas J. and {Atek}, Hakim and {Coe}, Dan and {Cutler}, Sam E. and {Dayal}, Pratika and {van Dokkum}, Pieter and {Feldmann}, Robert and {F{\"o}rster Schreiber}, Natascha M. and {Fujimoto}, Seiji and {Geha}, Marla and {Glazebrook}, Karl and {de Graaff}, Anna and {Greene}, Jenny E. and {Juneau}, St{\'e}phanie and {Kassin}, Susan and {Kriek}, Mariska and {Khullar}, Gourav and {Maseda}, Michael and {Mowla}, Lamiya A. and {Muzzin}, Adam and {Nanayakkara}, Themiya and {Nelson}, Erica J. and {Oesch}, Pascal A. and {Pacifici}, Camilla and {Pan}, Richard and {Papovich}, Casey and {Setton}, David J. and {Shapley}, Alice E. and {Smit}, Renske and {Stefanon}, Mauro and {Taylor}, Edward N. and {Williams}, Christina C.},
        title = "{The JWST UNCOVER Treasury Survey: Ultradeep NIRSpec and NIRCam Observations before the Epoch of Reionization}",
      journal = {\apj},
         year = 2024,
        month = oct,
       volume = {974},
       number = {1},
          eid = {92},
        pages = {92},
          doi = {10.3847/1538-4357/ad66cf},
archivePrefix = {arXiv},
       eprint = {2212.04026},
 primaryClass = {astro-ph.GA},
       adsurl = {https://ui.adsabs.harvard.edu/abs/2024ApJ...974...92B}
}

@article{brammer2023msaexp,
  title={msaexp: NIRSpec analyis tools},
  author={Brammer, Gabriel},
  journal={Zenodo},
  year={2023}
}

@article{brammer2008eazy,
  title={EAZY: a fast, public photometric redshift code},
  author={Brammer, Gabriel B and Van Dokkum, Pieter G and Coppi, Paolo},
  journal={The Astrophysical Journal},
  volume={686},
  number={2},
  pages={1503},
  year={2008},
  publisher={IOP Publishing}
}

@ARTICLE{2014MNRAS.442..900N,
       author = {{Nakajima}, Kimihiko and {Ouchi}, Masami},
        title = "{Ionization state of inter-stellar medium in galaxies: evolution, SFR-M$_{*}$-Z dependence, and ionizing photon escape}",
      journal = {\mnras},
         year = 2014,
        month = jul,
       volume = {442},
       number = {1},
        pages = {900-916},
          doi = {10.1093/mnras/stu902},
archivePrefix = {arXiv},
       eprint = {1309.0207},
 primaryClass = {astro-ph.CO},
       adsurl = {https://ui.adsabs.harvard.edu/abs/2014MNRAS.442..900N}
}

@article{suzuki2016iii,
  title={[O iii] emission line as a tracer of star-forming galaxies at high redshifts: comparison between H$\alpha$ and [O iii] emitters at z= 2.23 in HiZELS},
  author={Suzuki, TL and Kodama, T and Sobral, D and Khostovan, AA and Hayashi, M and Shimakawa, R and Koyama, Y and Tadaki, K-i and Tanaka, I and Minowa, Y and others},
  journal={Monthly Notices of the Royal Astronomical Society},
  volume={462},
  number={1},
  pages={181--189},
  year={2016},
  publisher={The Royal Astronomical Society}
}

@article{wold2025uncovering,
  title={UNCOVERing the Faint End of the z~ 7 [O iii] Luminosity Function with JWST’s F410M Medium Bandpass Filter},
  author={Wold, Isak GB and Malhotra, Sangeeta and Rhoads, James E and Weaver, John R and Wang, Bingjie},
  journal={The Astrophysical Journal},
  volume={980},
  number={2},
  pages={200},
  year={2025},
  publisher={IOP Publishing}
}

@article{laseter2024jades,
  title={JADES: Detecting [OIII] $\lambda$4363 emitters and testing strong line calibrations in the high-z Universe with ultra-deep JWST/NIRSpec spectroscopy up to z~ 9.5},
  author={Laseter, Isaac H and Maseda, Michael V and Curti, Mirko and Maiolino, Roberto and D’Eugenio, Francesco and Cameron, Alex J and Looser, Tobias J and Arribas, Santiago and Baker, William M and Bhatawdekar, Rachana and others},
  journal={Astronomy \& Astrophysics},
  volume={681},
  pages={A70},
  year={2024},
  publisher={EDP Sciences}
}

@ARTICLE{2013A&C.....3...23M,
       author = {{Murray}, S.~G. and {Power}, C. and {Robotham}, A.~S.~G.},
        title = "{HMFcalc: An online tool for calculating dark matter halo mass functions}",
      journal = {Astronomy and Computing},
         year = 2013,
        month = nov,
       volume = {3},
          eid = {23},
        pages = {23},
          doi = {10.1016/j.ascom.2013.11.001},
archivePrefix = {arXiv},
       eprint = {1306.6721},
 primaryClass = {astro-ph.CO},
       adsurl = {https://ui.adsabs.harvard.edu/abs/2013A&C.....3...23M}
}

@article{aghanim2020planck,
  title={Planck 2018 results. VI. Cosmological parameters},
  author={Aghanim, N and others},
  journal={Astron. Astrophys},
  volume={641},
  pages={A6},
  year={2020}
}

@article{sheth1999large,
  title={Large-scale bias and the peak background split},
  author={Sheth, Ravi K and Tormen, Giuseppe},
  journal={Monthly Notices of the Royal Astronomical Society},
  volume={308},
  number={1},
  pages={119--126},
  year={1999},
  publisher={The Royal Astronomical Society}
}

@article{reed2003evolution,
  title={Evolution of the mass function of dark matter haloes},
  author={Reed, Darren and Gardner, Jeffrey and Quinn, Thomas and Stadel, Joachim and Fardal, Mark and Lake, George and Governato, Fabio},
  journal={Monthly Notices of the Royal Astronomical Society},
  volume={346},
  number={2},
  pages={565--572},
  year={2003},
  publisher={Blackwell Science Ltd Oxford, UK}
}

@article{dekel2023efficient,
  title={Efficient formation of massive galaxies at cosmic dawn by feedback-free starbursts},
  author={Dekel, Avishai and Sarkar, Kartick C and Birnboim, Yuval and Mandelker, Nir and Li, Zhaozhou},
  journal={Monthly Notices of the Royal Astronomical Society},
  volume={523},
  number={3},
  pages={3201--3218},
  year={2023},
  publisher={Oxford University Press}
}

@article{li2024feedback,
  title={Feedback-free starbursts at cosmic dawn-Observable predictions for JWST},
  author={Li, Zhaozhou and Dekel, Avishai and Sarkar, Kartick C and Aung, Han and Giavalisco, Mauro and Mandelker, Nir and Tacchella, Sandro},
  journal={Astronomy \& Astrophysics},
  volume={690},
  pages={A108},
  year={2024},
  publisher={EDP Sciences}
}

@article{boyett2024massive,
  title={A massive interacting galaxy 510 million years after the Big Bang},
  author={Boyett, Kristan and Trenti, Michele and Leethochawalit, Nicha and Calabr{\'o}, Antonello and Metha, Benjamin and Roberts-Borsani, Guido and Dalmasso, Nicol{\'o} and Yang, Lilan and Santini, Paola and Treu, Tommaso and others},
  journal={Nature Astronomy},
  volume={8},
  number={5},
  pages={657--672},
  year={2024},
  publisher={Nature Publishing Group UK London}
}

@ARTICLE{2025arXiv250708245T,
       author = {{Tang}, Mengtao and {Stark}, Daniel P. and {Mason}, Charlotte A. and {Gelli}, Viola and {Chen}, Zuyi and {Topping}, Michael W.},
        title = "{The JWST Spectroscopic Properties of Galaxies at $z=9-14$}",
      journal = {arXiv e-prints},
         year = 2025,
        month = jul,
          eid = {arXiv:2507.08245},
        pages = {arXiv:2507.08245},
          doi = {10.48550/arXiv.2507.08245},
archivePrefix = {arXiv},
       eprint = {2507.08245},
 primaryClass = {astro-ph.GA},
       adsurl = {https://ui.adsabs.harvard.edu/abs/2025arXiv250708245T}
}

@article{fujimoto2024uncover,
  title={UNCOVER: a NIRSpec census of lensed galaxies at z= 8.50--13.08 probing a high-AGN fraction and ionized bubbles in the shadow},
  author={Fujimoto, Seiji and Wang, Bingjie and Weaver, John R and Kokorev, Vasily and Atek, Hakim and Bezanson, Rachel and Labbe, Ivo and Brammer, Gabriel and Greene, Jenny E and Chemerynska, Iryna and others},
  journal={The Astrophysical Journal},
  volume={977},
  number={2},
  pages={250},
  year={2024},
  publisher={IOP Publishing}
}

@article{sanders2023excitation,
  title={Excitation and ionization properties of star-forming galaxies at z= 2.0--9.3 with JWST/NIRSpec},
  author={Sanders, Ryan L and Shapley, Alice E and Topping, Michael W and Reddy, Naveen A and Brammer, Gabriel B},
  journal={The Astrophysical Journal},
  volume={955},
  number={1},
  pages={54},
  year={2023},
  publisher={IOP Publishing}
}

@article{witstok2021assessing,
  title={Assessing the sources of reionization: a spectroscopic case study of a 30$\times$ lensed galaxy at z~ 5 with Ly$\alpha$, C iv, Mg ii, and [Ne iii]},
  author={Witstok, Joris and Smit, Renske and Maiolino, Roberto and Curti, Mirko and Laporte, Nicolas and Massey, Richard and Richard, Johan and Swinbank, Mark},
  journal={Monthly Notices of the Royal Astronomical Society},
  volume={508},
  number={2},
  pages={1686--1700},
  year={2021},
  publisher={Oxford University Press}
}

@article{carnall2018inferring,
  title={Inferring the star formation histories of massive quiescent galaxies with BAGPIPES: evidence for multiple quenching mechanisms},
  author={Carnall, AC and McLure, RJ and Dunlop, JS and Dav{\'e}, R},
  journal={Monthly Notices of the Royal Astronomical Society},
  volume={480},
  number={4},
  pages={4379--4401},
  year={2018},
  publisher={Oxford University Press}
}

@ARTICLE{2025arXiv250701096G,
       author = {{Giovinazzo}, Emma and {Oesch}, Pascal A. and {Weibel}, Andrea and {Meyer}, Romain A. and {Witten}, Callum and {Bhagwat}, Aniket and {Brammer}, Gabriel and {Chisholm}, John and {de Graaff}, Anna and {Gottumukkala}, Rashmi and {Jecmen}, Michelle and {Katz}, Harley and {Leja}, Joel and {Marques-Chaves}, Rui and {Maseda}, Michael and {Shivaei}, Irene and {Trebitsch}, Maxime and {Verhamme}, Anne},
        title = "{Breaking Through the Cosmic Fog: JWST/NIRSpec Constraints on Ionizing Photon Escape in Reionization-Era Galaxies}",
      journal = {arXiv e-prints},
         year = 2025,
        month = jul,
          eid = {arXiv:2507.01096},
        pages = {arXiv:2507.01096},
          doi = {10.48550/arXiv.2507.01096},
archivePrefix = {arXiv},
       eprint = {2507.01096},
 primaryClass = {astro-ph.GA},
       adsurl = {https://ui.adsabs.harvard.edu/abs/2025arXiv250701096G}
}

@article{merlin2024astrodeep,
  title={ASTRODEEP-JWST: NIRCam-HST multi-band photometry and redshifts for half a million sources in six extragalactic deep fields},
  author={Merlin, E and Santini, P and Paris, D and Castellano, M and Fontana, A and Treu, T and Finkelstein, SL and Dunlop, JS and Haro, P Arrabal and Bagley, M and others},
  journal={Astronomy \& Astrophysics},
  volume={691},
  pages={A240},
  year={2024},
  publisher={EDP Sciences}
}

@article{fontana2000photometric,
  title={Photometric Redshifts and Selection of High-Redshift Galaxiesin the NTT and Hubble DeepFields},
  author={Fontana, Adriano and D’Odorico, Sandro and Poli, Francesco and Giallongo, Emanuele and Arnouts, Stephane and Cristiani, Stefano and Moorwood, Alan and Saracco, Paolo},
  journal={The Astronomical Journal},
  volume={120},
  number={5},
  pages={2206},
  year={2000},
  publisher={IOP Publishing}
}

@ARTICLE{2016MNRAS.462.1415C,
       author = {{Chevallard}, Jacopo and {Charlot}, St{\'e}phane},
        title = "{Modelling and interpreting spectral energy distributions of galaxies with BEAGLE}",
      journal = {\mnras},
         year = 2016,
        month = oct,
       volume = {462},
       number = {2},
        pages = {1415-1443},
          doi = {10.1093/mnras/stw1756},
archivePrefix = {arXiv},
       eprint = {1603.03037},
 primaryClass = {astro-ph.GA},
       adsurl = {https://ui.adsabs.harvard.edu/abs/2016MNRAS.462.1415C}
}

@ARTICLE{2000ApJ...533..682C,
       author = {{Calzetti}, Daniela and {Armus}, Lee and {Bohlin}, Ralph C. and {Kinney}, Anne L. and {Koornneef}, Jan and {Storchi-Bergmann}, Thaisa},
        title = "{The Dust Content and Opacity of Actively Star-forming Galaxies}",
      journal = {\apj},
         year = 2000,
        month = apr,
       volume = {533},
       number = {2},
        pages = {682-695},
          doi = {10.1086/308692},
archivePrefix = {arXiv},
       eprint = {astro-ph/9911459},
 primaryClass = {astro-ph},
       adsurl = {https://ui.adsabs.harvard.edu/abs/2000ApJ...533..682C}
}

@article{grudic2022dynamics,
  title={The dynamics and outcome of star formation with jets, radiation, winds, and supernovae in concert},
  author={Grudi{\'c}, Michael Y and Guszejnov, D{\'a}vid and Offner, Stella SR and Rosen, Anna L and Raju, Aman N and Faucher-Gigu{\`e}re, Claude-Andr{\'e} and Hopkins, Philip F},
  journal={Monthly Notices of the Royal Astronomical Society},
  volume={512},
  number={1},
  pages={216--232},
  year={2022},
  publisher={Oxford University Press}
}

@article{menon2022infrared,
  title={Infrared radiation feedback does not regulate star cluster formation},
  author={Menon, Shyam H and Federrath, Christoph and Krumholz, Mark R},
  journal={Monthly Notices of the Royal Astronomical Society},
  volume={517},
  number={1},
  pages={1313--1338},
  year={2022},
  publisher={Oxford University Press}
}

@article{bik2026spatially,
  title={Spatially resolved metallicity and ionization in the merging system Gz9p3 at z= 9.3},
  author={Bik, Arjan and {\'A}lvarez-M{\'a}rquez, Javier and G{\'o}mez, Alejandro Crespo and Colina, Luis and P{\'e}rez-Gonz{\'a}lez, Pablo G and {\"O}stlin, G{\"o}ran and Prieto, Carmen Blanco and Melinder, Jens and Langeroodi, Danial and Wright, Gillian and others},
  journal={arXiv preprint arXiv:2604.22460},
  year={2026}
}

@article{lee2024morphology,
  title={Morphology of Galaxies in JWST Fields: Initial Distribution and Evolution of Galaxy Morphology},
  author={Lee, Jeong Hwan and Park, Changbom and Hwang, Ho Seong and Kwon, Minseong},
  journal={The Astrophysical Journal},
  volume={966},
  number={1},
  pages={113},
  year={2024},
  publisher={The American Astronomical Society}
}

@article{somerville2026galaxy,
  title={Galaxy formation in the first billion years},
  author={Somerville, Rachel S},
  journal={arXiv preprint arXiv:2604.01445},
  year={2026}
}

\newpage

\appendix

\section{Physical properties of \uncovgal}\label{secA0}
\begin{figure*}[!htbp]
\setlength{\tabcolsep}{1pt}
\centering
\begin{tabular}{cccccc}
    \includegraphics[width=0.16\linewidth, height=2.5cm]{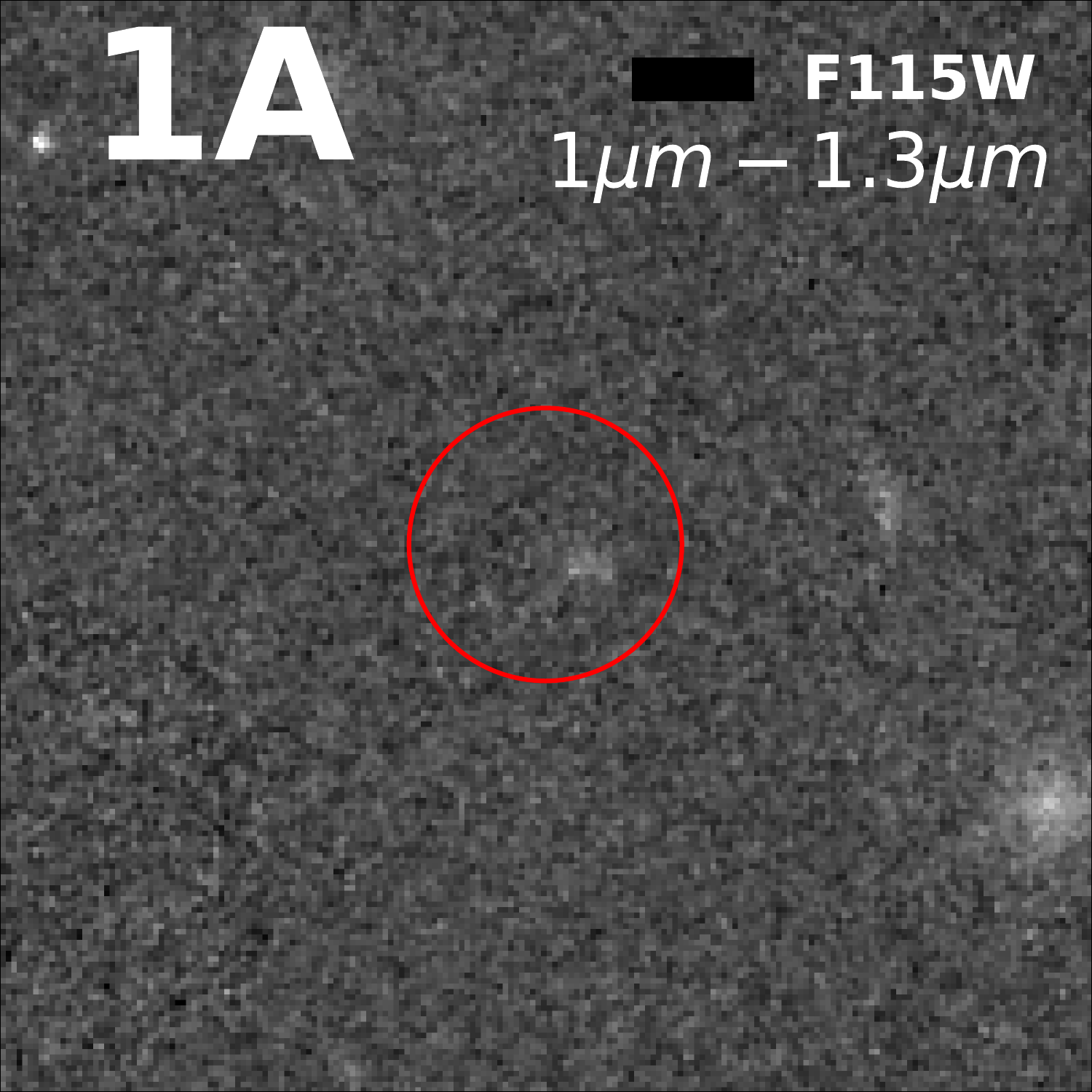} &
    \includegraphics[width=0.16\linewidth, height=2.5cm]{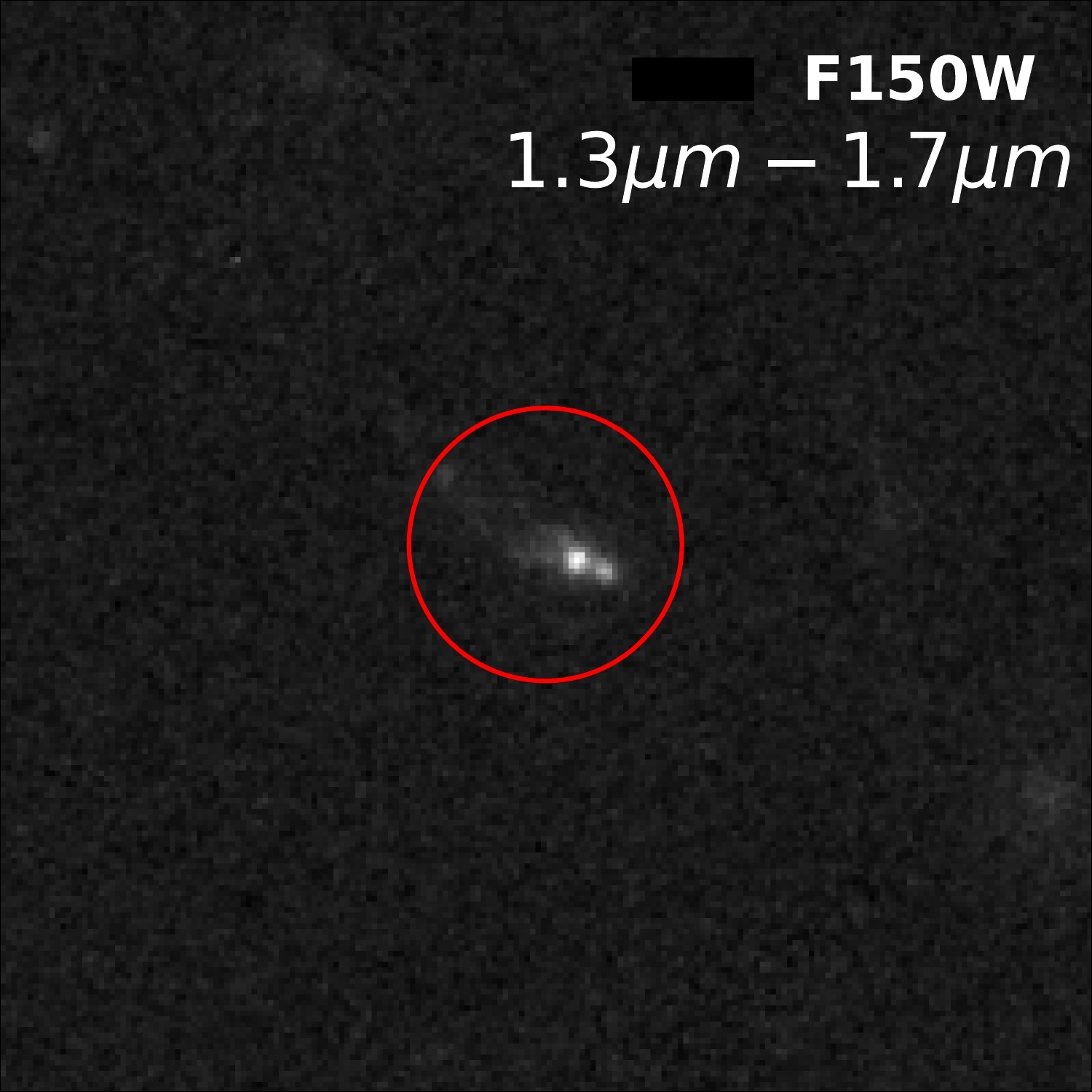} &
    \includegraphics[width=0.16\linewidth, height=2.5cm]{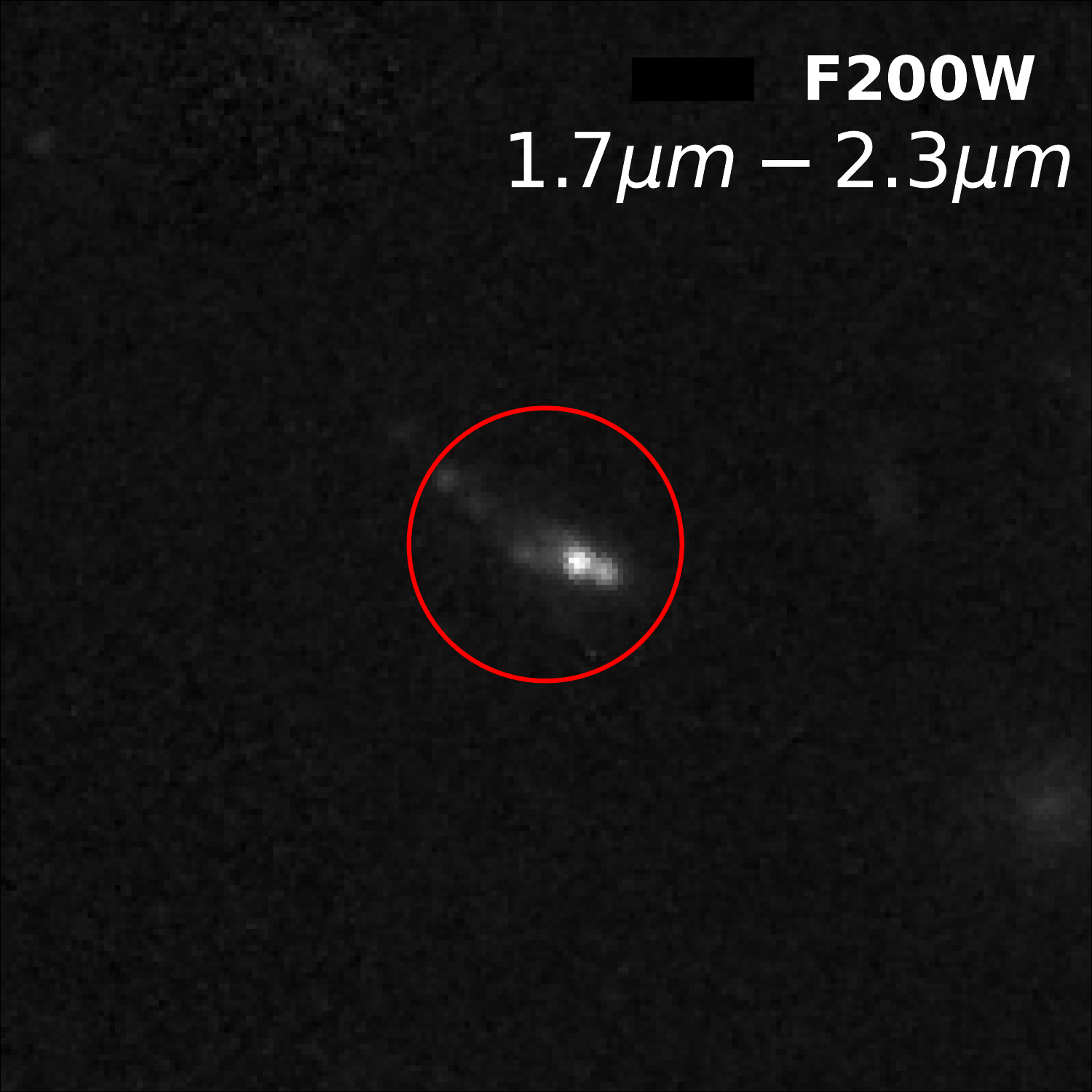} &
    \includegraphics[width=0.16\linewidth, height=2.5cm]{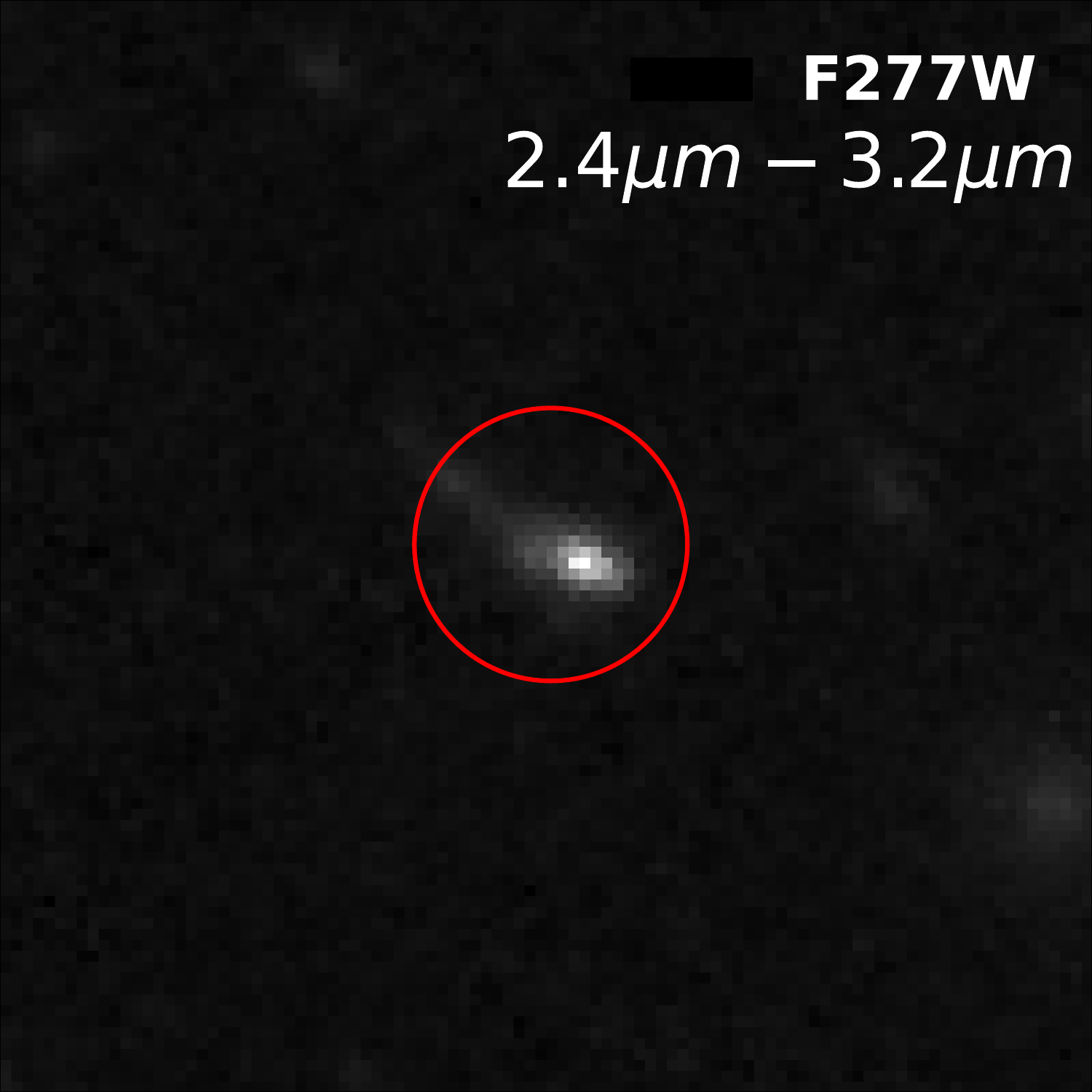} &
    \includegraphics[width=0.16\linewidth, height=2.5cm]{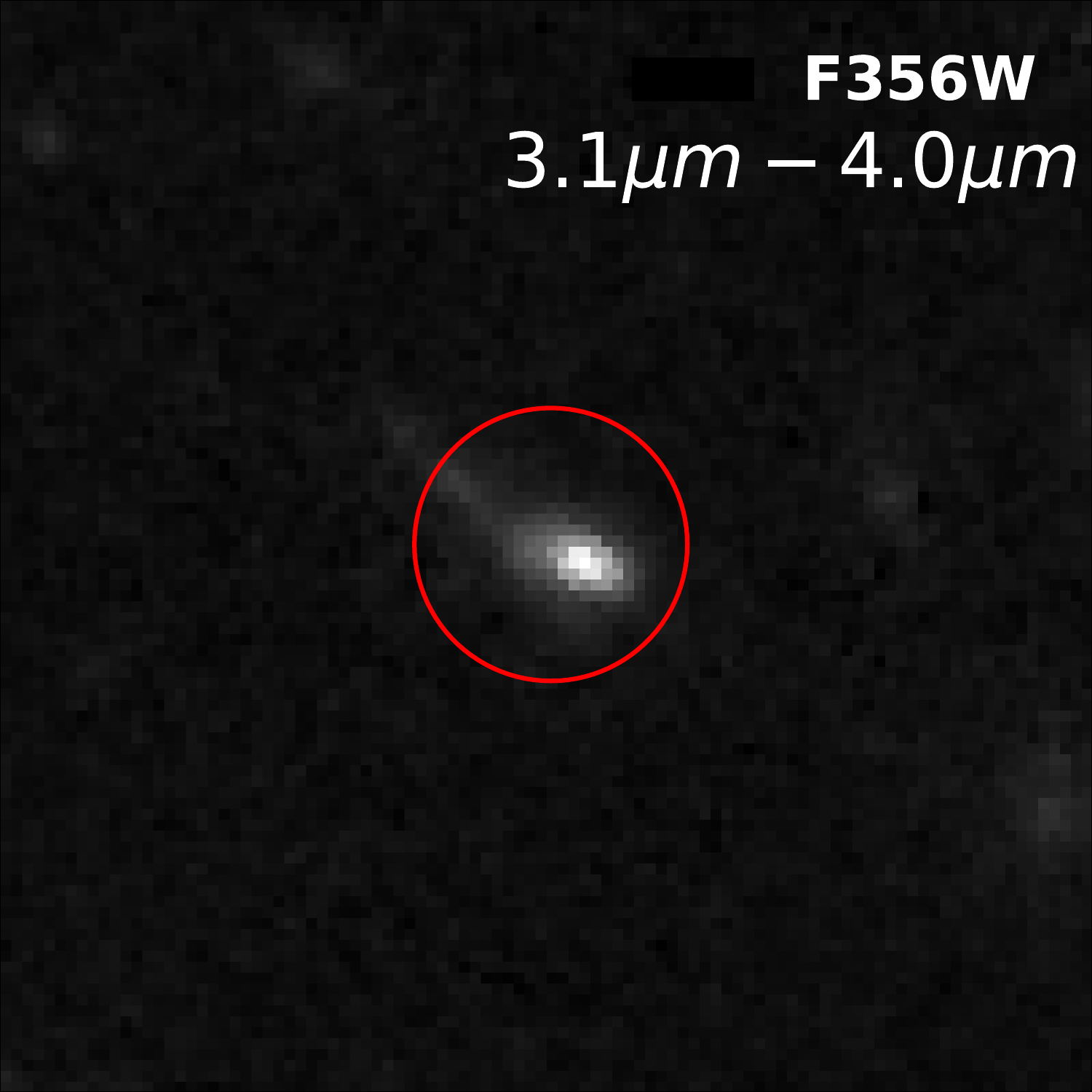} &
    \includegraphics[width=0.16\linewidth, height=2.5cm]{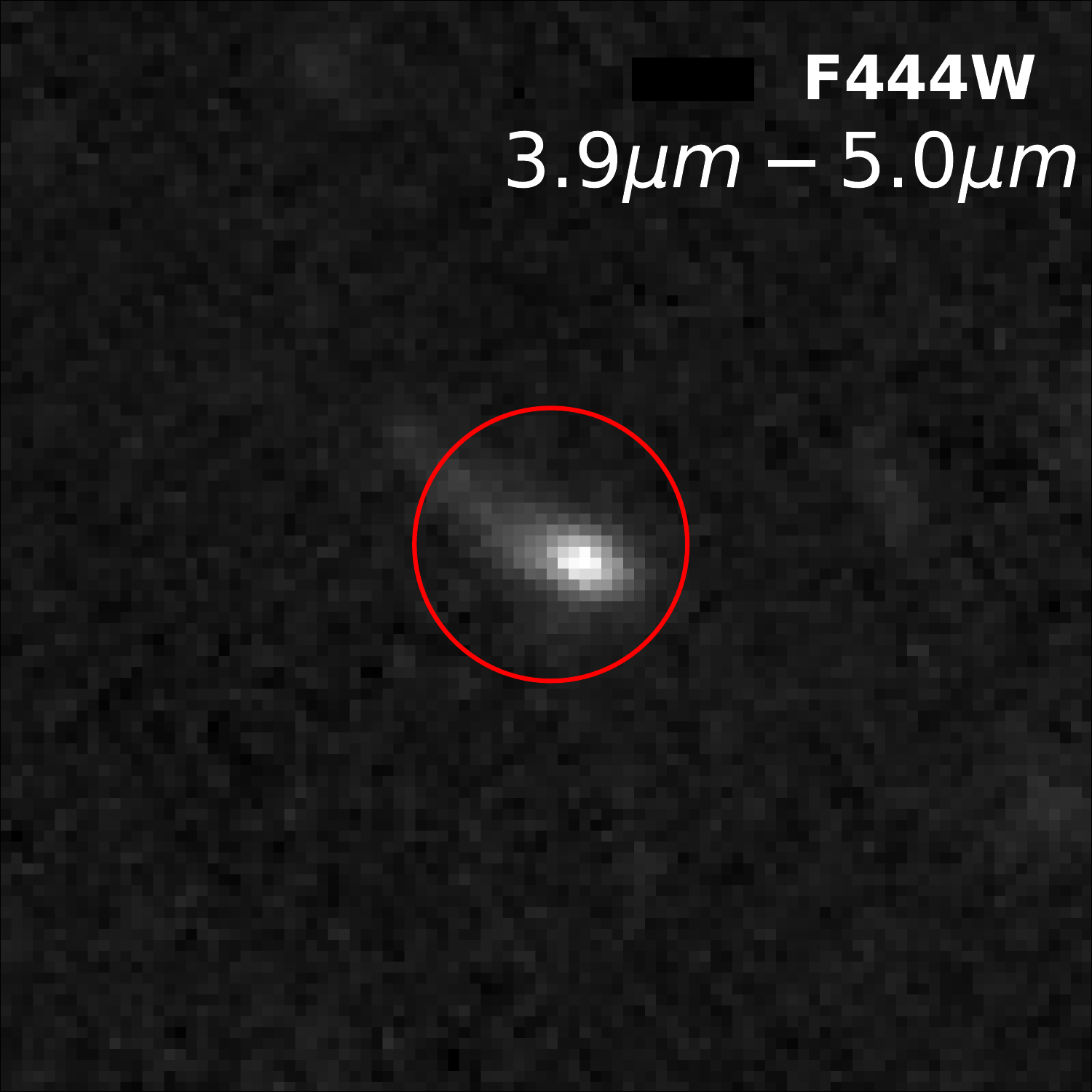}
\end{tabular}
\includegraphics[width=0.99\textwidth, height=7.0cm]{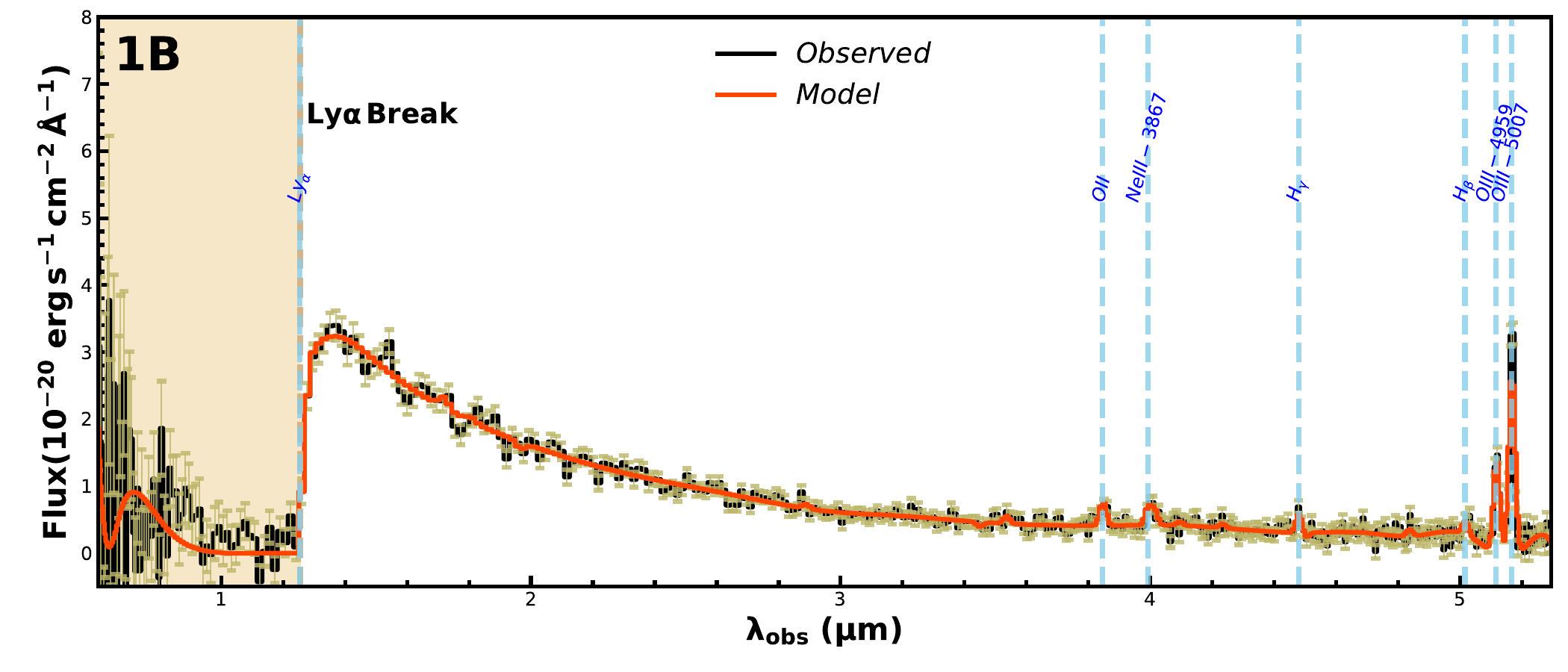}
\includegraphics[width=0.98\textwidth, height=4.0cm]{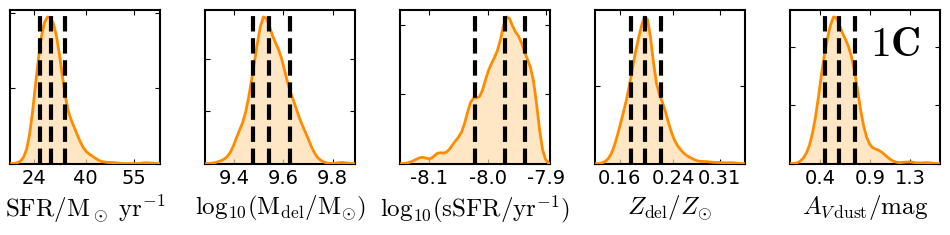}
\caption{\textbf{ Panel A}: The $4^{"} \times 4^{"}$, JWST postage stamps of \texttt{UNCOVER 3686} in various NIRCam filters \textbf{Panel B}: The $1D$ JWST/NIRSpec PRISM spectrum+model of \texttt{UNCOVER 3686}. \textbf{Panel C}: Posterior probability distributions of the physical parameters derived from SED fitting, including star formation rate, stellar mass, specific star formation rates, metallicity, and dust attenuation. Dashed vertical lines denote median values and the 16th–84th percentile confidence intervals. The subscript $'del'$ indicates the delayed star formation history model used for fitting the observed spectral energy density. }
\label{fig:spectrum}
\end{figure*}

Figure~\ref{fig:spectrum} summarizes the key observational
properties of \uncovgal. Panel~A shows the $4^{\prime\prime}
\times 4^{\prime\prime}$ JWST/NIRCam postage stamps in six
filters from F115W to F444W; the weak flux in F115W combined
with clear detections in redder bands is consistent with
Lyman-break dropout at $z>9$ \citep{atek2023revealing}.
Panel~B shows the 1D JWST/NIRSpec PRISM spectrum from
\cite{2025ApJ...982...51P}, which unambiguously exhibits a
strong Lyman break and prominent [O\,{\sc iii}]
$\lambda\lambda4959,5007$ emission lines confirming
$z_{\rm spec}=9.31^{+0.01}_{-0.01}$. Panel~C shows the
posterior probability distributions of the key physical
parameters derived from the joint photometric and
spectroscopic SED fitting with \texttt{BAGPIPES}
\citep{carnall2018inferring}, including star formation
rate, stellar mass, specific star formation rate,
metallicity, and dust attenuation.

\section{Morphology of \uncovgal}\label{secA3}
\begin{figure*}
    \centering
    \includegraphics[width=1.0\textwidth]{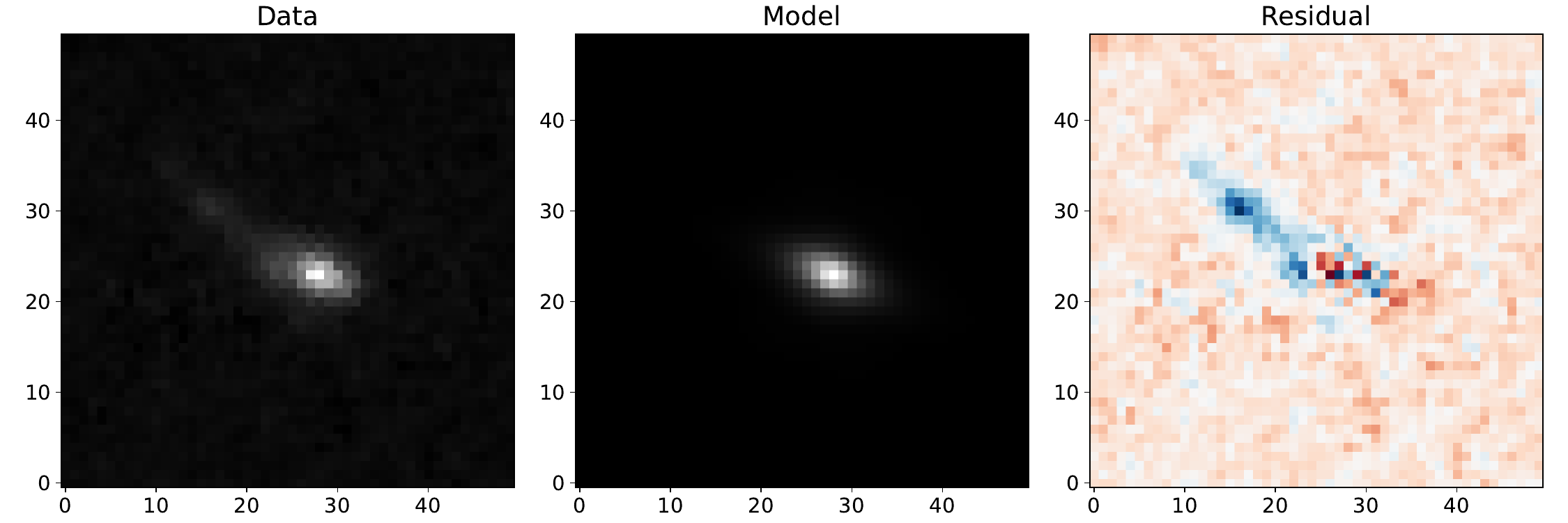}
    \caption{We show the morphological measurement of \texttt{UNCOVER 3686} performed by fitting a single S\'{e}rsic index to the galaxy image in the F277W band. Background-subtracted image (left), PSF-convolved single Sérsic model of the central component (middle), and residual map (right). The S\'{e}rsic fit describes the main stellar body, while the residuals reveal an extended asymmetric tail.}
    \label{appfig:morphology}
\end{figure*}

We model the surface brightness distribution of \texttt{UNCOVER 3686} using a two-dimensional S\'{e}rsic profile convolved with the instrumental point spread function (PSF). We first extract a cutout centered on the galaxy coordinates from the science image using the world coordinate system. The local background level and noise properties are estimated using a median background model, and we subtract this background from the image before further analysis. We identify sources in the cutout using a segmentation map constructed with a detection threshold of $5\sigma$ above the background noise. The segment corresponding to the target galaxy is then used to construct a mask that isolates the galaxy while excluding neighboring sources. We estimate the galaxy centroid and obtain an initial guess for the effective radius from flux-weighted moments and cumulative aperture photometry. We then fit a two-dimensional S\'{e}rsic model characterized by central intensity $I_e$, effective radius $R_e$, S\'{e}rsic index $n$, axis ratio $q$, and position angle to the masked image. The model is convolved with the PSF using Fourier convolution to account for the instrumental response. The best-fitting parameters are obtained by minimizing the noise-weighted residuals between the data and the model using a bounded least-squares optimization. To quantify the structures not captured by the S\'{e}rsic model, we analyze the residual image obtained after subtracting the best-fit model. We identify statistically significant residual emission using a $3\sigma$ threshold relative to the local RMS noise, excluding pixels within one effective radius to avoid contamination from the main galaxy light. The residual mask is mildly dilated to recover faint low surface brightness emission, and small noise islands are removed. We define the spatial extent of the extended tail as the maximum projected distance between the galaxy center and the farthest significant residual pixel. We show the results from the morphological fitting in Figure \ref{appfig:morphology}.

\section{Spectral modeling of \uncovgal}\label{secA1}
We fitted the available \emph{JWST} NIRCam+HST photometry together with the 1D NIRSpec PRISM spectrum of \texttt{UNCOVER 3686} using the \texttt{BAGPIPES} spectral fitting code \citep{carnall2018inferring}. We use photometric measurements from a combination of Hubble Space Telescope (ACS/WFC F606W, F814W; WFC3/IR F105W, F125W, F160W) and James Webb Space Telescope NIRCam (F090W, F115W, F150W, F200W, F210M, F277W, F356W, F444W) filters,  and use the spectra of \uncovgal from \cite{2025ApJ...982...51P} We confirm the reliability of the source by checking that the \texttt{use\_phot} and \texttt{flag\_kron} are set to $1$ and $0$, respectively \citep{2024ApJS..270....7W}. We ensure that the quality flags \texttt{flag\_successful\_spectrum} is set to 1, and \texttt{flag\_zspec\_qual} is set to 3, indicating that the target spectrum was successfully observed and reduced, and that the redshifts were detected securely from at least two or more spectral features. The spectroscopic redshift is derived using the \texttt{msaexp} pipeline \citep{brammer2023msaexp}, which determines the best-fit redshift by minimizing the $\chi^{2}$ between the observed spectrum and a set of templates. The fitting is performed using the physically motivated \texttt{blue\_sfhz\_13} template set from the \texttt{EAzY} code package \citep{brammer2008eazy}. The \texttt{blue\_sfhz\_13} template forbids unphysical star formation histories (SFHs) that are older than the age of the Universe.

\begin{figure*}
    \resizebox{1.0\textwidth}{!}{%
        \includegraphics{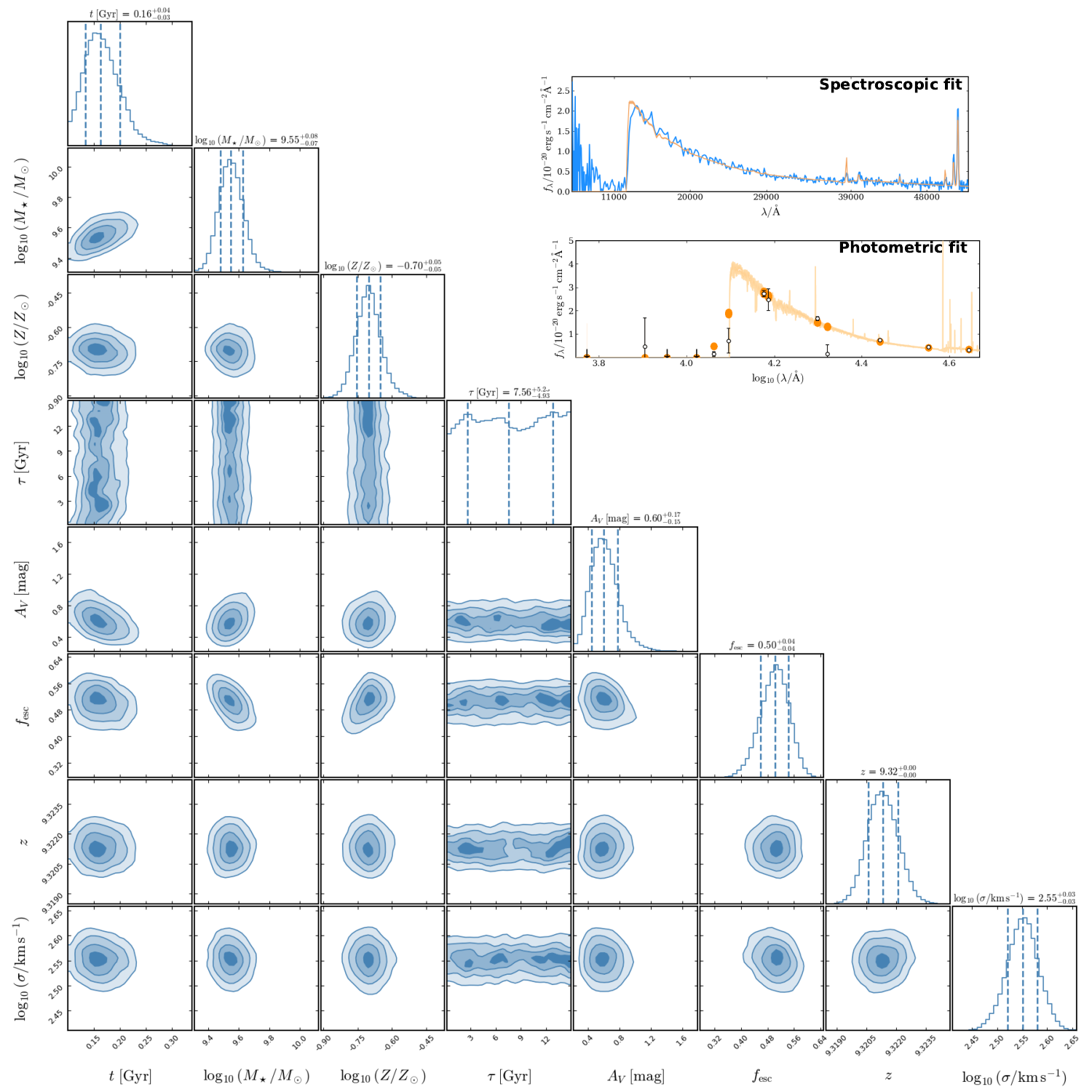}}
    \caption{Corner plot showing the posterior distributions of key physical parameters inferred from the joint photometric and
    spectroscopic SED fitting using \texttt{BAGPIPES}. The vertical dashed lines indicate the median and 16th/84th percentiles.
    Inset panels show the posterior median best-fit models compared to the observed photometric and spectroscopic data.}
    \label{appfig:corner-plot}
\end{figure*}

Our model describes the stellar emission, nebular emission, dust attenuation, kinematic broadening, and spectrophotometric calibration of the data, and the posterior distributions were sampled using nested sampling. The stellar component is modeled using the 2016 updated version of the \texttt{BC03} \citep{2016MNRAS.462.1415C} stellar population synthesis models and a delayed-$\tau$ star formation history of the form $\mathrm{SFR}(t) \propto t\,\exp(-t/\tau)$, allowing the stellar age to vary between $0.1$ and $13.8$ Gyr, the star-formation timescale $\tau$ between $0.03$ and $15$ Gyr, the total stellar mass formed over $1 < \log_{10}(M_{\star}/M_{\odot}) < 15$, and the stellar metallicity over $10^{-4} < Z_{\star}/Z_{\odot} < 2.5$ with a logarithmic prior. Nebular emission is included self-consistently with the ionization parameter fixed to $\log U=-2.19$ as inferred from the observed [Ne\,III]/[O\,II] ratio, while the escape fraction of ionizing photons is allowed to vary between $0.001$ and $1.0$. Dust attenuation is modeled using the \cite{2000ApJ...533..682C} attenuation law with the $V$-band attenuation allowed to vary over $0 < A_V < 5$ mag. The redshift is allowed to vary with a Gaussian prior centered at $z=9.3$ with $\sigma=0.05$, and the model spectrum is convolved with a Gaussian velocity kernel with velocity dispersion allowed to vary between $1$ and $10^{4}\,\mathrm{km\,s^{-1}}$ using a logarithmic prior. To account for residual spectrophotometric calibration uncertainties, we include a second-order Bayesian polynomial correction to the spectrum, and we also fit an additional white-noise scaling parameter that rescales the pipeline-provided spectral uncertainties. To place \texttt{UNCOVER 3686} in the context of the galaxy population at similar redshifts, we derive stellar masses for galaxies in the \texttt{ASTRODEEP-JWST} catalog by fitting their NIRCam photometry in F090W, F115W, F150W, F200W, F277W, F356W, F410M, and F444W filters with \texttt{BAGPIPES}. We adopt the same stellar population models, star-formation history parameterization, and priors described above. In this case, the redshifts are fixed to the catalog photometric redshift values. These fits are used solely to estimate stellar masses for the galaxy population employed in the abundance matching analysis. The resulting posterior distributions of the physical parameters are shown in Figure \ref{appfig:corner-plot} and summarized in Table \ref{tab:sed_params}.

\begin{table*}
\label{apptab:summary}
\centering
\caption{ Summary of SED-fitting parameters, priors, and posterior constraints
derived from the joint photometric and spectroscopic fitting. Quoted uncertainties
correspond to the 16th, 50th, and 84th percentiles of the posterior distributions.}
\label{tab:sed_params}
\begin{tabular}{lccccc}
\hline\hline
Parameter & Prior range & Prior type & 16th & 50th & 84th \\
\hline
\multicolumn{6}{c}{\textbf{Star-formation history (Delayed-$\tau$ model)}} \\
\hline
Age [Gyr]                      & $0.1$--$13.8$      & Uniform        & 0.134 & 0.163 & 0.200 \\
$\tau$ [Gyr]                   & $0.03$--$15$     & Uniform        & 2.636 & 7.564 & 12.792 \\
$\log_{10}(M_\star/M_\odot)$   & $1$--$15$        & Uniform        & 9.478 & 9.549 & 9.625 \\
$Z_\star/Z_\odot$              & $10^{-4}$--$2.5$ & Logarithmic   & 0.176 & 0.199 & 0.223 \\
\hline
\multicolumn{6}{c}{\textbf{Dust attenuation}} \\
\hline
$A_V$ [mag]                    & $0$--$5$         & Uniform        & 0.450 & 0.603 & 0.776 \\
\hline
\multicolumn{6}{c}{\textbf{Nebular emission}} \\
\hline
$f_{\rm esc}$                  & $0.001$--$1.0$   & Uniform        & 0.459 & 0.503 & 0.543 \\
$\log U$                       & $-2.19$ (fixed)  & --             & --    & --    & --    \\
\hline
\multicolumn{6}{c}{\textbf{Kinematics and redshift}} \\
\hline
Redshift $z$                   & $0$--$10$        & Gaussian       & 9.321 & 9.321 & 9.322 \\
$\sigma_{\rm v}$ [km\,s$^{-1}$]& $1$--$10^4$      & Logarithmic   & 331.7 & 356.2 & 380.7 \\
\hline
\multicolumn{6}{c}{\textbf{Spectral calibration}} \\
\hline
$P_0$                          & $0.5$--$2.5$     & Gaussian       & 1.606 & 1.651 & 1.701 \\
$P_1$                          & $-0.5$--$0.5$    & Gaussian       & $-0.177$ & $-0.102$ & $-0.028$ \\
$P_2$                          & $-0.5$--$0.5$    & Gaussian       & 0.196 & 0.237 & 0.278 \\
\hline
\multicolumn{6}{c}{\textbf{Noise model}} \\
\hline
$a_{\rm noise}$                & $1$--$10$        & Logarithmic   & 1.298 & 1.345 & 1.391 \\
\hline\hline
\end{tabular}
\end{table*}

\section{Abundance matching}\label{secA2}
We compute the efficiencies at fixed cumulative halo abundances using the \texttt{hmf} package \citep{2013A&C.....3...23M}. We use the present-day Hubble constant, $H_{0}=67.32$ \kmps \mpcinv; the $z=0$ density parameter for matter, $\Omega_{m}=0.3158$, the slope of the primordial power spectrum of density fluctuations, $n_{s}=0.96605$; and the cosmic baryon fraction, $f_{b}=\Omega_{b}/\Omega_{m}=0.156$ \citep{aghanim2020planck}. The comoving number density of halos above a mass $M_{h}$ at a redshift $z$ is given by
\begin{equation}
n(>M_{h},z) = \int^{\infty}_{M_{h}} dM   \frac{dn(M,z)}{dM}. 
\label{eq:n_Mh}
\end{equation}

We use the halo mass function, $dn(M_h,z)/dM_h$, from \cite{sheth1999large}. 
Since every galaxy is expected to be associated with a dark matter halo, the number density of galaxies with $M_{\star}$ or greater must be equal to the number density of halos that are massive enough to host them. Therefore, the abundance matching requires
\begin{equation}
  n_{\rm gal}(>M_\star,z) = n_h (> M_h,z) \,
\end{equation}
which further provides the value of $M_h$ corresponding to a given $M_\star$. The SFE is then 
\begin{equation}
\varepsilon = M_\star/(f_b M_h)
\end{equation}
However, note that the Sheth-Tormen halo mass function typically overestimates the abundance of the massive halos between $M_h=\pow{10-11}$ \msun by nearly $20\% - 50\%$ at $z\approx 10$ \citep{boylan2023stress,reed2003evolution}, thereby providing a conservative lower limit on the estimated SFE.

Each galaxy is detected with a completeness fraction $C_j \in (0,1]$, evaluated from sigmoid fits to the \texttt{ASTRODEEP-JWST} completeness curves \citep{merlin2024astrodeep} at 
its survey field. The observed cumulative number density is corrected as
\begin{equation}
n_{\rm gal}(>M_\star^{(i)}, z) =
\frac{1}{V_{\rm survey}}
\sum_{j=1}^{i} \frac{1}{C_j},
\end{equation}
\noindent where each galaxy is upweighted by $1/C_j$ to account for undetected counterparts. Cosmic variance is quantified per galaxy using \texttt{galcv} \citep{trapp2020flexible} at the matched halo mass, survey area, and redshift, yielding a fractional uncertainty $\sigma_{\rm cv}$. The perturbed number densities,
\begin{equation}
n_\pm = n_{\rm gal}(1 \pm \sigma_{\rm cv}),    
\end{equation}
are independently matched to the halo mass function to obtain $\varepsilon_\pm$. The total uncertainty on $\varepsilon$ combines the stellar mass and cosmic variance contributions in quadrature:
\begin{equation}
\delta\varepsilon =
\sqrt{
\delta\varepsilon_{M_\star}^2 +
\delta\varepsilon_{\rm cv}^2}
\end{equation}

\end{document}